\documentclass[11pt]{article}

\usepackage{cite}
\usepackage{amsmath,amssymb,amsfonts}
\usepackage{graphicx}
\usepackage{amsthm}
\usepackage{textcomp}
\usepackage{tikz}
\usepackage{pgfplots}
\usepackage{caption}
\usepackage{subcaption}
\usepackage{xcolor}
\usepackage{algorithmicx}
\usepackage[noend]{algpseudocode}
\usepackage{algorithm}
\usepackage{mathtools}
\usepackage{hyperref}
\allowdisplaybreaks 

\newtheorem{definition}{Definition} 
\newtheorem{theorem}{Theorem} 
\newtheorem{assumption}{Assumption} 
\newtheorem{problem}{Problem} 
\newtheorem{remark}{Remark}
\newtheorem{lemma}{Lemma}
\newtheorem{corollary}{Corollary}
\newtheorem{example}{Example}
\usepackage{array}
\usepackage{authblk}
\usepackage{setspace}
\usepackage{fullpage,etoolbox}

\title{Reactive and Risk-Aware Control for Signal Temporal Logic\thanks{This work was supported in part by the Swedish Research Council (VR), the European Research Council (ERC), the Swedish Foundation for Strategic Research (SSF),  the EU H2020 Co4Robots project, the Knut and Alice Wallenberg Foundation (KAW), the DARPA Assured Autonomy program, and the AFOSR grant FA9550-19-1-0265 (Assured Autonomy in Contested Environments).} }
\author[1]{Lars Lindemann}
\author[1]{George J. Pappas}
\author[2]{Dimos V. Dimarogonas}
\affil[1]{Department of Electrical and Systems Engineering, University of Pennsylvania}
\affil[2]{Division of Decision and Control Systems,  KTH Royal Institute of Technology}

\begin{document}

\maketitle

\begin{abstract}
The deployment of autonomous systems in uncertain and dynamic environments has raised fundamental  questions. Addressing these is pivotal to build fully autonomous systems and requires a systematic integration of planning and control. We first propose reactive risk signal interval temporal logic (ReRiSITL) as an extension of signal temporal logic (STL) to formulate complex spatiotemporal specifications. Unlike STL, ReRiSITL allows to consider uncontrollable propositions that may model humans as well as random environmental events such as sensor failures. Additionally, ReRiSITL allows to incorporate risk measures, such as (but not limited to) the Conditional Value-at-Risk, to measure the risk of violating certain spatial specifications.  Second, we propose an algorithm to check if an ReRiSITL specification is satisfiable. For this purpose, we abstract the ReRiSITL specification into a timed signal transducer and devise a game-based approach. Third, we propose a  reactive planning and control framework for dynamical control systems under ReRiSITL specifications. 
\end{abstract}

\section{Introduction}
\label{sec:introduction}

Temporal  logics allow to express temporal properties in a logical framework providing an expressive specification language. \emph{Signal temporal logic} (STL)  is a predicate-based temporal logic that offers  many appealing advantages~\cite{maler2004monitoring}. In particular, STL allows to impose quantitative temporal properties, e.g., combinations of surveillance (``visit regions A, B, and C every $10-60$ sec"), safety (``always between $5-25$ sec stay at least $1$ m away from D"), and many others. Indeed, there is a rich body of literature on the control of dynamical systems under STL specifications, e.g., \cite{raman1,pant2018fly,lindemann2018control}.

However, a key obstacle to deploying such control frameworks in real-world settings is to account for uncertain and dynamic environments. In particular, objects of interests may be estimated by simultaneous localization and mapping algorithms and  be described as probability distributions, see e.g., \cite{yiannis_iros} and \cite{fu2016optimal}, so that one may want to consider risk. Also, random events such as sensor failures or humans requesting assistance play an increasing  role. While there has been  recent work  addressing some of these challenges, e.g., \cite{farahani2018shrinking,safaoui2020control,SadighRSS16,gundana2020event}, there exists no  reactive and risk-aware planning and control framework with formal correctness guarantees.  We claim that no one has rigorously addressed the reactive planning problem for systems under STL specifications. Towards addressing this shortcoming, we leverage ideas from formal methods, risk theory, control theory, game theory, and timed automata theory.

\subsection{Related Work} For the control under STL specifications,  mixed integer linear programs \cite{raman1,sadraddini2015robust,belta2019formal} have been presented that encode the STL specification at hand.  Nonconvex optimization programs \cite{pant2018fly,mehdipour} and reinforcement learning approaches \cite{varnai2019prescribed,muniraj2018enforcing} have further been proposed and particularly use the quantitative semantics associated with an STL specification \cite{fainekos2009robustness}. A timed automata-based planning framework has been presented in our previous work \cite{lindemann2019efficient} where we decompose the STL specification into STL subspecifications. Feedback control laws that implement such STL subspecifications, which are timed transitions, have appeared in \cite{lindemann2018control,verginis2018timed,fiaz2019fast,garg2019control,yang2020continuous,ramasubramanian2019linear,niu2020control,mavridis2019robot}.

Linear temporal logic (LTL)  is a proposition-based temporal logic,  less expressive than STL, that allows to impose qualitative temporal properties. Existing control approaches leverage automata-based synthesis \cite{kloetzer2008fully,kantaros2018sampling,sahin2017synchronous}.  Metric interval temporal logic (MITL) is a proposition-based temporal logic with quantitative temporal properties \cite{alur1996benefits}, hence more expressive than LTL but less expressive than STL. An MITL specification can be translated into a language equivalent timed automaton \cite{alur1996benefits}. If the accepted language of this automaton is not empty \cite{alur1994theory}, the MITL specification is satisfiable.  For point-wise MITL semantics,  a tool to perform this translation has been presented in \cite{brihaye2017m}. Point-wise semantics, however, do not guarantee the satisfaction of the MITL specification in continuous time. The procedure of \cite{alur1996benefits}, for continuous-time semantics, is complex and not compositional. The results from \cite{maler2006mitl,ferrere2019real} are more intuitive and present a compositional way to construct a timed signal transducer for an MITL specification. The authors in \cite{maler1995synthesis} have proposed a way to control timed automata by reformulating it as a timed two player game, played between controllable (the system) and uncontrollable (environment) events, see also \cite{asarin1998controller,asarin1994symbolic,asarin1999soon}.
 
The underlying assumption in these previous works is that the environment is perfectly known. For LTL, this assumption has been relaxed in \cite{yiannis_iros,lahijanian2016iterative,fu2016optimal}. Specifically, \cite{yiannis_iros} and \cite{fu2016optimal} assume that the environment is modeled as a semantic map. Target beliefs in surveillance games and Markov decisions  process-based approaches are  presented in \cite{bharadwaj2018synthesis} and \cite{guo2018probabilistic}. Probabilistic computational tree logic and  distribution temporal logic \cite{vasile2016control} account for state distributions and can take chance constraints into account,  but only consider qualitative temporal properties and do not consider risk measures \cite{rockafellar2000optimization,majumdar2020should}. The works in \cite{piterman2006synthesis} and \cite{kress2009temporal} consider the generalized reactivity(1) fragment, which explicitly accounts for  dynamic environments.  For STL, the works in \cite{farahani2018shrinking} and \cite{SadighRSS16} consider chance constraints, whereas \cite{safaoui2020control} and \cite{lindemann2020control} already incorporate risk measures without, however, considering random environmental events. Such events have been considered for STL in \cite{gundana2020event}. The proposed reactive control strategy in  \cite{gundana2020event} has been evaluated empirically, but without providing formal guarantees. A reactive counter-example guided framework was proposed in \cite{raman2015reactive} where, however, the risk of violating  certain spatial specifications is not considered. Furthermore, only bounded specifications are considered while the STL specification is not allowed to explicitly depend on the environment.

\subsection{Contributions} In this paper, our \emph{first contribution} is to propose reactive risk signal interval temporal logic (ReRiSITL). Compared with STL,  ReRiSITL has two distinct features and hence generalizes STL. First, ReRiSITL specifications may contain uncontrollable propositions that allow to model humans, or in general other agents, and environmental events such as sensor failures or communication dropouts.  Second, ReRiSITL allows to incorporate  risk measures by considering risk predicates so that the risk of violating certain spatial specifications can be taken into account. Such risk predicates can take different risk measures into account, as for instance the conditional value-at-risk (CVaR). Our \emph{second contribution} is an algorithm that allows to check if such an ReRiSITL specification is satisfiable. To do so, we abstract the ReRiSITL specification into a timed signal transducer using and adapting the results from \cite{ferrere2019real} and then following a game-based strategy similarly to \cite{maler1995synthesis}. The \emph{third contribution} is a planning and control framework for dynamical control systems under ReRiSITL specifications. The main elements here are a well defined timed abstraction of the control system that relies on existing feedback control laws as presented in \cite{verginis2018timed,fiaz2019fast,garg2019control,yang2020continuous,ramasubramanian2019linear,niu2020control,mavridis2019robot}. We then propose to use a combination of a game-based approach, graph search techniques, and  replanning. We  remark that our approach is, to the best of our knowledge, the first to incorporate past temporal operators and we hereby establish a connection between monitoring and reactive control.



%
%

\textbf{Structure.} Section \ref{sec:backgound} presents  ReRiSITL and the problem formulation. Section \ref{sec:strategy} presents the algorithm to check if an ReRiSITL specification is satisfiable. Sections  \ref{sec:determinization} and \ref{sec:planning}  propose the planning and control framework for dynamical control systems under ReRiSITL specifications. Simulations and conclusions are provided in Sections \ref{sec:simulations} and  \ref{sec:conclusion}.

\section{Preliminaries and Problem Formulation}
\label{sec:backgound}

True and false are encoded as $\top:=\infty$ and $\bot:=-\infty$ with $\mathbb{B}:=\{\top,\bot\}$. Let $\mathbb{R}$, $\mathbb{Q}$, and $\mathbb{N}$ be the real, rational, and natural numbers, respectively, while $\mathbb{R}_{\ge 0}$ ($\mathbb{R}_{> 0}$) and $\mathbb{Q}_{\ge 0}$ ($\mathbb{Q}_{> 0}$) denote their respective nonnegative (positive) subsets. For  $t\in\mathbb{R}_{\ge 0}$ and $I\subseteq\mathbb{R}_{\ge 0}$, let  $t\oplus I$ and $t\ominus I$ denote the Minkowski sum and the Minkowski difference of $t$ and $I$, respectively. For two sets $\mathcal{X}$ and $\mathcal{Y}$, we use the notation  $\mathcal{F}(\mathcal{X},\mathcal{Y})$ to denote the set of all measurable functions that map from  $\mathcal{X}$ to $\mathcal{Y}$. An element $f\in \mathcal{F}(X,Y)$ is hence a function $f:\mathcal{X}\to \mathcal{Y}$. 

 Let $(\Omega,\mathcal{B}_\Omega,P_\Omega)$ be a probability space where $\Omega$ is the sample space, $\mathcal{B}_\Omega$ is the Borel $\sigma$-algebra of $\Omega$, and $P_\Omega:\mathcal{B}_\Omega\to[0,1]$ is a probability measure. A vector of random variables is a measurable function $\boldsymbol{X}:\Omega \to \mathbb{R}^{\tilde{n}}$  defined on a probability space $(\Omega,\mathcal{B}_\Omega,P_\Omega)$ where $\tilde{n}\in\mathbb{N}$. We can associate the probability space $(\mathbb{R}^{\tilde{n}},\mathcal{B}_{\mathbb{R}^{\tilde{n}}},P_{\boldsymbol{X}})$ with $\boldsymbol{X}$ with probability measure  $P_{\boldsymbol{X}}:\mathcal{B}_{\mathbb{R}^{\tilde{n}}}\to[0,1]$ defined as 
\begin{align*}
P_{\boldsymbol{X}}(B):=P_\Omega(\boldsymbol{X}^{-1}(B))
\end{align*} 
for  Borel sets $B\in\mathcal{B}_{\mathbb{R}^{\tilde{n}}}$  and where $\boldsymbol{X}^{-1}(B):=\{\omega\in\Omega|\boldsymbol{X}(\omega)\in B\}$ is the inverse image. Let  $\tilde{\boldsymbol{\mu}}:=EV[\boldsymbol{X}]$ and $\tilde{\Sigma}$  be the expected value  and covariance matrix of $\boldsymbol{X}$, respectively, while $\mathcal{N}(\tilde{\boldsymbol{\mu}},\tilde{\Sigma})$ denotes the multivariate normal distribution. We remark that  all important symbols that have been or will be introduced in this paper are summarized in Table \ref{tab:111}.

\begin{table*}
	\centering
	\begin{tabular}{| m{3.25cm} | m{13cm} | } 
		\hline
		\footnotesize{Symbol} & \footnotesize{Meaning}  \\
		\hline
		\hline
		\footnotesize{$\mathcal{F}(\mathcal{X},\mathcal{Y})$} & \footnotesize{Set of all measurable functions mapping from a set $\mathcal{X}$ into a set $\mathcal{Y}$.} \\
		\hline
		\footnotesize{$\boldsymbol{x}$, $\boldsymbol{s}$} & \footnotesize{The function $\boldsymbol{x}:\mathbb{R}_{\ge 0}\to \mathbb{R}^n$ denotes a deterministic signal, while the element $\boldsymbol{s}\in\mathcal{F}(\mathbb{R}_{\ge 0},\mathbb{B}^{|M^\text{uc}|})$ denotes a random signal.} \\
		\hline
		\footnotesize{$\boldsymbol{X}$, $\tilde{\boldsymbol{\mu}},\tilde{\Sigma}$} & \footnotesize{The function $\boldsymbol{X}:\Omega\to\mathbb{R}^{\tilde{n}}$ denotes a random variable with expected value $\tilde{\boldsymbol{\mu}}\in\mathbb{R}^{\tilde{n}}$ and covariance matrix $\tilde{\Sigma}\in\mathbb{R}^{\tilde{n}\times \tilde{n}}$.} \\
		\hline
		\footnotesize{$h$} & \footnotesize{The function $h:\mathbb{R}^n\times\mathbb{R}^{\tilde{n}}$ denotes predicate functions.} \\
		\hline
		\footnotesize{$M^\text{Ri}$, $M^\text{uc}$, $M$} & \footnotesize{$M^\text{Ri}$: set of risk predicates, $M^\text{uc}$: set of uncontrollable propositions, $M$: set of risk predicates and uncontrollable propositions.} \\
		\hline
		\footnotesize{$\mu^\text{Ri}$, $\mu^\text{uc}$} & \footnotesize{The element $\mu^\text{Ri}\in M^\text{Ri}$ is a risk predicate, while the element $\mu^\text{uc}\in M^\text{uc}$ is an uncontrollable proposition.} \\
		\hline
		\footnotesize{$R$, $\beta$, $\gamma$, } & \footnotesize{The function $R:\mathcal{F}(\Omega,\mathbb{R})\to\mathbb{R}$ denotes a risk measure, $\beta$ is a risk level, and $\gamma$ is a risk threshold.} \\
		\hline
		\footnotesize{$(\boldsymbol{x},\boldsymbol{s},\boldsymbol{X},t)\models \phi$} & \footnotesize{Semantics of an ReRiSITL specification $\phi$ indicating that $\boldsymbol{x}$, $\boldsymbol{s}$, and $\boldsymbol{X}$ satisfy $\phi$ at time $t$.} \\
		\hline
		\footnotesize{$AP$} & \footnotesize{Set of (atomic) propositions for MITL specifications.} \\
		\hline
		\footnotesize{$BC$} & \footnotesize{The function $BC$, e.g., applied as $BC(AP)$, denotes the set of all Boolean combinations (negations, conjunctions, disjunctions) over $AP$.  } \\
		\hline
		\footnotesize{$Tr$, $Tr^{-1}$} & \footnotesize{The transformation $\varphi=Tr(\phi)$ transforms an ReRiSITL specification $\phi$ into an MITL specification $\varphi$; $Tr^{-1}$ is the inverse. } \\
		\hline
		\footnotesize{$TST_\varphi$, $TST_\phi$} & \footnotesize{Timed signal transducers for the MITL specification $\varphi$ and the ReRiSITL specification $\phi$.} \\
		\hline
		\footnotesize{$RA$, $RA_C$, $\overline{RA}_C$} & \footnotesize{The functions $RA$, $RA_C$, $\overline{RA}_C$, e.g., applied as $RA(TST_\phi)$,  are different versions of the region automaton of  $TST_\phi$. } \\
		\hline
		\footnotesize{$d_p$, $d_\mu$} & \footnotesize{The plan $d_p:\mathbb{R}_{\ge 0}\to BC(AP)$ is constructed for a specification $\phi$, $d_\mu$ is simply its projection to  $M$ via $Tr^{-1}$.} \\
		\hline
		\footnotesize{$\pi$, $\hat{\pi}$, $W$} & \footnotesize{The functions $\pi$ and $\hat{\pi}$ are different versions of the controllable predecessor for Algorithm \ref{alg:1} providing the winning condition $W$.} \\
		\hline
		\footnotesize{$\mathfrak{X}_m$, $\mathfrak{X}_m^{\text{EV}}$, $\mathfrak{X}_m^{\text{VaR}}$, $\mathfrak{X}_m^{\text{CVaR}}$} & \footnotesize{The sets $\mathfrak{X}_m^{\text{EV}}$, $\mathfrak{X}_m^{\text{VaR}}$, $\mathfrak{X}_m^{\text{CVaR}}$ are risk constrained sets that are determinized into the set $\mathfrak{X}_m$.} \\
		\hline
		\footnotesize{$\mu^\text{det}$, $M^\text{det}$, $\hat{M}$} & \footnotesize{The element $\mu^\text{det}\in M^\text{det}$ is a deterministic predicate; $\hat{M}$ is the set of deterministic predicates and uncontrollable propositions.} \\
		\hline
		\footnotesize{$TST_\theta$, $TST_\theta^\text{m}$} & \footnotesize{Timed signal transducers for the ReSITL specification $\theta$ and the product automaton.} \\
		\hline
	\end{tabular}
	\caption{Summary of the most important notation used throughout the paper.}
	\label{tab:111}
\end{table*}

\subsection{Reactive Risk Signal Interval Temporal Logic }
\label{sec:ReRiSTL}

To define \emph{reactive risk signal interval temporal logic} (ReRiSITL), let  
\begin{align*}
h:\mathbb{R}^n\times \mathbb{R}^{\tilde{n}}\to\mathbb{R}
\end{align*}
 be a measurable function,  referred to as the \emph{predicate function}, where $n,\tilde{n}\in\mathbb{N}$. Let  
 \begin{align*}
\boldsymbol{x}:\mathbb{R}_{\ge 0}\to\mathbb{R}^n
 \end{align*}
  be a deterministic signal and let 
  \begin{align*}
\boldsymbol{X}:\Omega \to \mathbb{R}^{\tilde{n}}
  \end{align*}
   be a vector of random variables defined on the probability space $(\Omega,\mathcal{B}_\Omega,P_\Omega)$.\footnote{We remark that $\boldsymbol{X}$ can be assumed to be a stochastic process $\boldsymbol{X}(t)$. To avoid further technical complexity, this is not followed in this paper.} At time $t$, the probability space $(\mathbb{R},\mathcal{B}_\mathbb{R},P_h)$ can be associated with  $h(\boldsymbol{x}(t),\boldsymbol{X})$, a random variable, where $P_h$ is derived from the probability space $(\mathbb{R}^{\tilde{n}},\mathcal{B}_{\mathbb{R}^{\tilde{n}}},P_{\boldsymbol{X}})$.


We consider \emph{risk predicates} for ReRiSITL based on risk measures as advocated in \cite{rockafellar2000optimization,majumdar2020should} towards an axiomatic risk assessment. A \emph{risk measure} 
\begin{align*}
R:\mathcal{F}(\Omega,\mathbb{R})\to\mathbb{R}
\end{align*} 
allows to exclude behavior which is deemed more risky than other behavior.  We are interested in $R(-h(\boldsymbol{x}(t),\boldsymbol{X}))$ to argue about the risk of violating $h(\boldsymbol{x}(t),\boldsymbol{X})\ge 0$. The truth value of a risk predicate $\mu^{\text{Ri}}:\mathbb{R}^n\times \mathbb{R}^{\tilde{n}}\to\mathbb{B}$ at time $t$  is obtained as
\begin{align}\label{eq:risk_predicate}
\mu^{\text{Ri}}(\boldsymbol{x}(t),\boldsymbol{X})&:=\begin{cases}
\top & \text{if } R(-h(\boldsymbol{x}(t),\boldsymbol{X}))\le \gamma\\
\bot &\text{otherwise }
\end{cases}
\end{align} 
for a risk threshold $\gamma\in\mathbb{R}$. There are various choices of $R(\cdot)$, see  \cite{majumdar2020should} for an overview. We consider the expected value (EV), the Value-at-Risk (VaR), and the Conditional Value-at-Risk (CVaR).  The expected value of $-h(\boldsymbol{x}(t),\boldsymbol{X})$, denoted by $EV[-h(\boldsymbol{x}(t),\boldsymbol{X})]$, provides a risk neutral risk measure. More risk averse  are the VaR and the  CVaR as in \cite{rockafellar2000optimization}. The VaR of $-h(\boldsymbol{x}(t),\boldsymbol{X})$ for $\beta\in(0,1)$ is defined as
\begin{align*}
VaR_\beta(-h(\boldsymbol{x}(t),\boldsymbol{X}))&:=\min(d\in\mathbb{R}|P_h(-h(\boldsymbol{x}(t),\boldsymbol{X})\le d)\ge \beta),
\end{align*}
i.e.,  the worst case  $1-\beta$ probability quantile. 
\begin{remark}\label{rem:prbb}
	Note that $VaR_\beta(-h(\boldsymbol{x}(t),\boldsymbol{X}))\le \gamma$ is equivalent to $P_h(-h(\boldsymbol{x}(t),\boldsymbol{X})\le \gamma)\ge \beta$ so that our framework includes chance constraints as for instance used in \cite{SadighRSS16}. 
\end{remark}

The CVaR of $-h(\boldsymbol{x}(t),\boldsymbol{X})$ for a risk level $\beta$ is given by
\begin{align*}
CVaR_\beta(-h(\boldsymbol{x}(t),\boldsymbol{X}))&:=EV[-h(\boldsymbol{x}(t),\boldsymbol{X}))|-h(\boldsymbol{x}(t),\boldsymbol{X}))>VaR_\beta(-h(\boldsymbol{x}(t),\boldsymbol{X}))],
\end{align*}
i.e., the conditional expected value of $-h(\boldsymbol{x}(t),\boldsymbol{X})$ relative to $-h(\boldsymbol{x}(t),\boldsymbol{X})$ being greater than or equal to the VaR. Let now  $M^{\text{Ri}}$ denote a set of  risk predicates.

Let  $M^{\text{uc}}$ be a set of \emph{uncontrollable propositions} $\mu^\text{uc}$ and $\boldsymbol{s}\in \mathcal{F}(\mathbb{R}_{\ge 0}, \mathbb{B}^{|M^{\text{uc}}|})$ be a random Boolean signal corresponding to the truth values of the propositions in $M^{\text{uc}}$ over time.\footnote{The proposition $\mu^{\text{uc}}$ is labeled uncontrollable because $\boldsymbol{s}$ is assumed to be a random signal generated by an unknown underlying stochastic process, as highlighted by the notation $\boldsymbol{s}\in \mathcal{F}(\mathbb{R}_{\ge 0}, \mathbb{B}^{|M^{\text{uc}}|})$.} Define also the projection of $\boldsymbol{s}$ onto  $\mu^{\text{uc}}\in M^{\text{uc}}$ as proj$_{\mu^{\text{uc}}}(\boldsymbol{s}):\mathbb{R}_{\ge 0}\to \mathbb{B}$, i.e., the truth value of $\mu^{\text{uc}}$ over time. 

Define the set of risk predicates and uncontrollable propositions as
\begin{align*}
M:=M^{\text{Ri}} \cup M^{\text{uc}}.
\end{align*}
For $\mu\in M$, the syntax of ReRiSITL is now given as
\begin{align}\label{eq:full_ReRiSTL}
\phi \; ::= \; \top \; | \; \mu \; | \;  \neg \phi \; | \; \phi' \wedge \phi'' \; | \; \phi'  U_I \phi'' \; | \; \phi' \underline{U}_I \phi''
\end{align}
where $\phi'$ and $\phi''$ are ReRiSITL formulas and where $U_I$ and $\underline{U}_I$ are the future and past until operators.  We restrict the time interval $I$ to belong to the nonnegative rationals, i.e., $I\subseteq\mathbb{Q}_{\ge 0}$. Additionally,  we require that  $I$  is not a singleton, i.e., $I$ is not allowed to be of the form $I:=[a,a]$ for $a\in\mathbb{Q}_{\ge 0}$. Note that the former assumption is not restrictive, while  the latter excludes punctuality constraints. We remark that these assumptions are commonly made  \cite{alur1996benefits}. Also define
\begin{align*}
\phi' \vee \phi''&:=\neg(\neg\phi' \wedge \neg\phi'') & \text{(disjunction),}\\
F_I\phi&:=\top U_I \phi & \text{(future eventually),}\\
\underline{F}_I\phi&:=\top \underline{U}_I \phi & \text{(past eventually),}\\
G_I\phi&:=\neg F_I\neg \phi & \text{(future always),}\\
\underline{G}_I\phi&:=\neg \underline{F}_I\neg \phi & \text{(past always)}.
\end{align*}
We say that an ReRiSITL formula $\phi$ is in positive normal form if no negation occurs within $\phi$ \cite{sadraddini2015robust}. Let $(\boldsymbol{x},\boldsymbol{s},\boldsymbol{X},t)\models \phi$ denote the satisfaction relation as defined next.  

\begin{definition}[ReRiSITL Semantics]\label{def:ReRiSITL}
	We recursively define the continuous-time semantics of ReRiSITL  as 
	\begin{align*}
	&(\boldsymbol{x},\boldsymbol{s},\boldsymbol{X},t) \models
	\mu^{\text{Ri}}&\text{ iff }\;&R(-h(\boldsymbol{x}(t),\boldsymbol{X}))\le \gamma,\\
	&(\boldsymbol{x},\boldsymbol{s},\boldsymbol{X},t) \models
	\mu^{\text{uc}} &\text{ iff }\; &\text{proj}_{\mu^{\text{uc}}}(\boldsymbol{s})(t)=\top,\\ 
	&(\boldsymbol{x},\boldsymbol{s},\boldsymbol{X},t) \models \neg\phi&\text{ iff }\;&\neg((\boldsymbol{x},\boldsymbol{s},\boldsymbol{X},t) \models \phi),\\ 
	&(\boldsymbol{x},\boldsymbol{s},\boldsymbol{X},t) \models \phi' \wedge \phi''&\text{ iff }\;&(\boldsymbol{x},\boldsymbol{s},\boldsymbol{X},t) \models \phi' \wedge (\boldsymbol{x},\boldsymbol{s},\boldsymbol{X},t) \models \phi'',\\ &(\boldsymbol{x},\boldsymbol{s},\boldsymbol{X},t) \models \phi' U_I \phi'' &\text{ iff }\;&\exists t'' \in t\oplus I \text{ such that } (\boldsymbol{x},\boldsymbol{s},\boldsymbol{X},t'')\models \phi''\wedge \forall t'\in (t,t'') \text{,}(\boldsymbol{x},\boldsymbol{s},\boldsymbol{X},t') \models \phi',\\
&(\boldsymbol{x},\boldsymbol{s},\boldsymbol{X},t) \models \phi' \underline{U}_I \phi'' &\text{ iff }\;&\exists t'' \in t\ominus I \text{ such that } (\boldsymbol{x},\boldsymbol{s},\boldsymbol{X},t'')\models \phi''\wedge \forall t'\in (t,t'') \text{,}(\boldsymbol{x},\boldsymbol{s},\boldsymbol{X},t') \models \phi'. 
	\end{align*}
\end{definition}
\begin{remark} Quantitative semantics can be defined similarly to \cite{fainekos2009robustness}  to determine how well $\boldsymbol{x}$, $\boldsymbol{s}$, and $\boldsymbol{X}$ satisfy $\phi$ at time $t$. Here, one has to consider $\gamma-R(-h(\boldsymbol{x}(t),\boldsymbol{X}))$ and $\text{proj}_{\mu^{\text{uc}}}(\boldsymbol{s})(t)$ for risk predicates and uncontrollable propositions, and  recursively apply the operations from \cite[Def. 10]{fainekos2009robustness}.
\end{remark}

\begin{example}\label{ex:2}
	Consider the workspace  in Fig. \ref{fig:sim_overview} with regions R1, R2, O1, and O2 described by a normal distribution 
	\begin{align*}
	\boldsymbol{X}:=\begin{bmatrix}\boldsymbol{X}_\text{R1}^T & \boldsymbol{X}_\text{R2}^T & \boldsymbol{X}_\text{O1}^T & \boldsymbol{X}_\text{O2}^T \end{bmatrix}^T\sim \mathcal{N}(\tilde{\boldsymbol{\mu}},\tilde{\Sigma})
	\end{align*} 
	with expected value and covariance according to
	\begin{align*}
	\tilde{\boldsymbol{\mu}}&:=\begin{bmatrix}8&8&2&4& 5&7&5&5\end{bmatrix}^T\\
	\tilde{\Sigma}&:=\text{diag}(0.1,0.1,0.1,0.1,0.1,0.1,0.1,0.1).
	\end{align*} 
	Consider also the
	following ReRiSITL specification 
	\begin{align*}
	\phi&:= F_{(0,5)} \mu_\text{R1}^\text{Ri}\wedge G_{[0,\infty)}\Big(\mu_\text{O1}^\text{Ri}\wedge \mu_\text{O2}^\text{Ri} \wedge \big(\underline{F}_{(0,1)}\mu^\text{uc}\implies F_{(0,3)}\mu_\text{R2}^\text{Ri}\big)\Big)
	\end{align*}
	where $\mu_\text{R1}^\text{Ri}$ and $\mu_\text{R2}^\text{Ri}$ encode the probability of reaching  the regions R1 and R2 using the VaR, $\mu_\text{O1}^\text{Ri}$ and $\mu_\text{O2}^\text{Ri}$ encode the risk of colliding with obstacles O1 and O2 using the CVaR, and $\mu^\text{uc}$ is an uncontrollable proposition. The specification $\phi$ encodes to reach R1 within $5$ time units with probability of at least $0.8$, while always having a risk of colliding with obstacles O1 and O2 lower than $0$. Furthermore, whenever the uncontrollable proposition $\mu^\text{uc}$, e.g., encoding a human requesting assistance,  was true within the last $1$ time unit, it should follow that R2 is reached within $3$ time units with probability $0.8$. We emphasize the use of the past operator $\underline{F}_{(0,1)}$ in $\phi$ that specifies a form of reactive monitoring. In particular, the predicate functions are 
	\begin{align*}
h_\text{R1}^\text{Ri}(\boldsymbol{x}(t),\boldsymbol{X}):=\epsilon-\|\boldsymbol{x}(t)-\boldsymbol{X}_\text{R1}\|^2\\
h_\text{R2}^\text{Ri}(\boldsymbol{x}(t),\boldsymbol{X}):=\epsilon-\|\boldsymbol{x}(t)-\boldsymbol{X}_\text{R2}\|^2
	\end{align*}
	where $\epsilon:=0.5$ and $R_\text{R1}()$ and $R_\text{R2}()$ encode the VaR with $\beta_\text{R1}=\beta_\text{R2}:=0.8$ and $\gamma_\text{R1}=\gamma_\text{R2}:=0$. Recall that, according to Remark \ref{rem:prbb}, the risk predicate $\mu_\text{R1}^\text{Ri}$ using VaR encodes the probability that $h_\text{R1}^\text{Ri}(\boldsymbol{x}(t),\boldsymbol{X})\ge 0$ is greater than $0.8$. Let also 
	\begin{align*}
	h_\text{O1}^\text{Ri}(\boldsymbol{x}(t),\boldsymbol{X}):=\|\boldsymbol{x}(t)-\boldsymbol{X}_\text{O1}\|^2-\epsilon\\
	h_\text{O2}^\text{Ri}(\boldsymbol{x}(t),\boldsymbol{X}):=\|\boldsymbol{x}(t)-\boldsymbol{X}_\text{O2}\|^2-\epsilon
	\end{align*}
	where the risk measures $R_\text{O1}()$ and $R_\text{O2}()$ encode the CVaR with $\beta_\text{O1}=\beta_\text{O2}:=0.9$ and $\gamma_\text{O1}=\gamma_\text{O2}:=0$. 
				\begin{figure}
		\centering
		\includegraphics[scale=0.5]{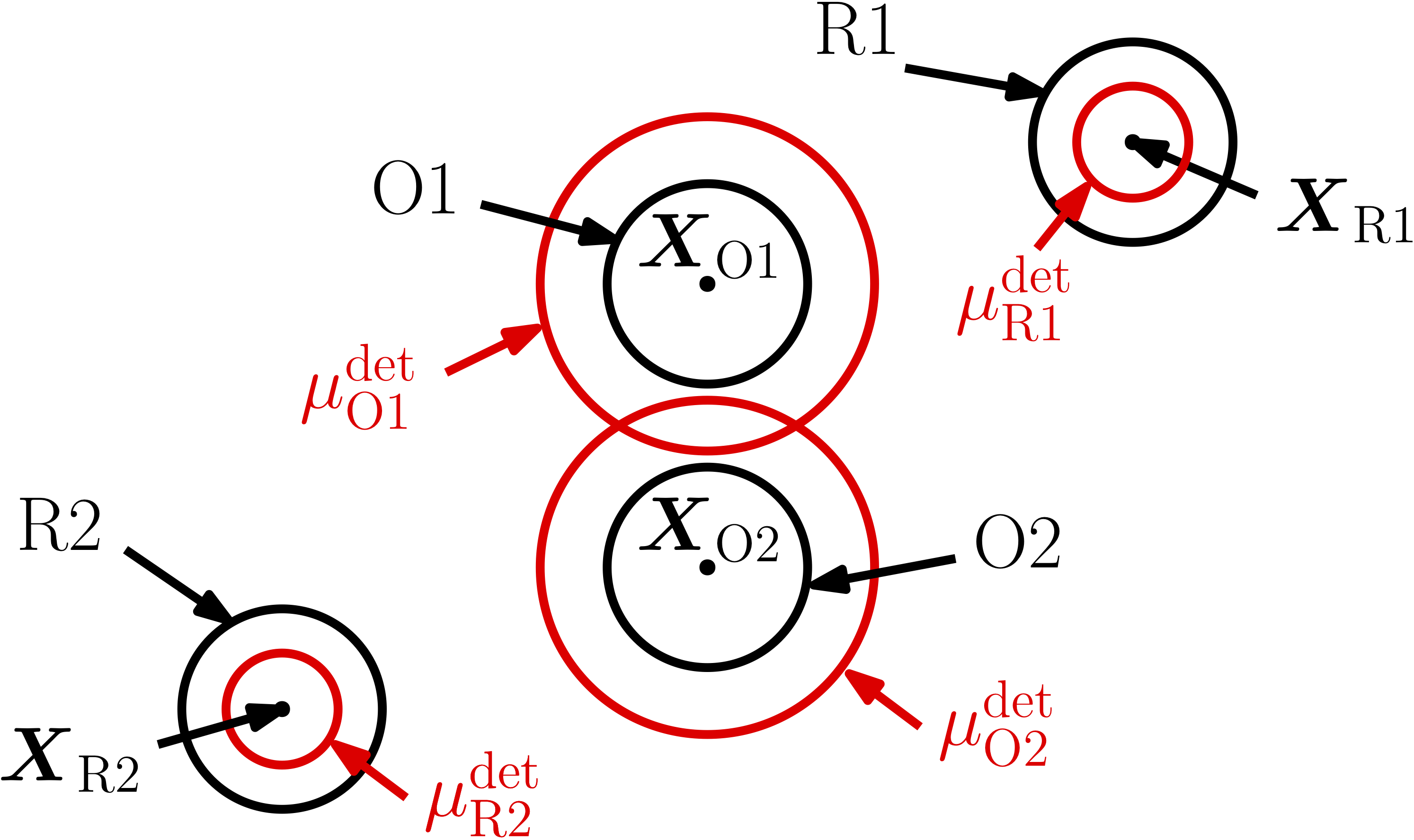}\caption{Overview of the  workspace in Example \ref{ex:2}.}\label{fig:sim_overview}
	\end{figure}
\end{example}

To define satisfiability of an ReRiSITL specification, we need to take into account that propositions in $M^{\text{uc}}$ are uncontrollable. We first define what a \emph{nonanticipative strategy} is. A strategy 
\begin{align*}
\boldsymbol{x}_\text{na}:\mathcal{F}(\mathbb{R}_{\ge 0},\mathbb{B}^{|M^{\text{uc}}|})\to\mathcal{F}(\mathbb{R}_{\ge 0},\mathbb{R}^n)
\end{align*}
 is nonanticipative if: for any $t\ge 0$ and for any two signals $\boldsymbol{s},\boldsymbol{s}'\in \mathcal{F}(\mathbb{R}_{\ge 0},\mathbb{B}^{|M^{\text{uc}}|})$ with $\boldsymbol{s}(\tau)=\boldsymbol{s}'(\tau)$ for all $\tau\in[0,t]$, it holds that $\boldsymbol{x}_\text{na}(\boldsymbol{s})(\tau)=\boldsymbol{x}_\text{na}(\boldsymbol{s}')(\tau)$ for all $\tau\in[0,t]$. This means that $\boldsymbol{x}_\text{na}(\boldsymbol{s})$  takes, at time $t$, only current and past values  of $\boldsymbol{s}$ into account, i.e., $\boldsymbol{s}(\tau)$ where $\tau\le t$. This makes sense under the assumption that $\boldsymbol{s}(t)$ can only be observed at time~$t$.
\begin{definition}[ReRiSITL Satisfiability]\label{def:satisfiability}
	For a given $\boldsymbol{X}$, an ReRiSITL formula $\phi$ is said to be satisfiable if $\forall \boldsymbol{s}\in\mathcal{F}(\mathbb{R}_{\ge 0},\mathbb{B}^{|M^{\text{uc}}|})$, there exists a nonanticipative strategy $\boldsymbol{x}_\text{na}:\mathcal{F}(\mathbb{R}_{\ge 0},\mathbb{B}^{|M^{\text{uc}}|})\to\mathcal{F}(\mathbb{R}_{\ge 0},\mathbb{R}^n)$ s.t. $(\boldsymbol{x}_\text{na}(\boldsymbol{s}),\boldsymbol{s},\boldsymbol{X},0)\models \phi$. 
\end{definition}

Later in the paper, we will replace risk predicates by \emph{deterministic predicates} as originally used in STL.  For a given constant   $c\in\mathbb{R}$, the truth value of such a deterministic predicate $\mu^{\text{det}}:\mathbb{R}^n\times \mathbb{R}^{\tilde{n}}\to\mathbb{B}$ at time $t$ is obtained as
\begin{align}\label{eq:STL_predicate}
\mu^{\text{det}}(\boldsymbol{x}(t),\tilde{\boldsymbol{\mu}}):=\begin{cases}
\top & \text{if } h(\boldsymbol{x}(t),\tilde{\boldsymbol{\mu}})\ge c\\
\bot &\text{otherwise.}
\end{cases}
\end{align}
where we have replaced $\boldsymbol{X}$ in $h$ by its expected value  $\tilde{\boldsymbol{\mu}}$.

If now all risk predicates $\mu^{\text{Ri}}\in M^\text{Ri}$ are replaced by deterministic predicates  $\mu^{\text{det}}$, then $\phi$ is called a reactive signal interval temporal logic (ReSITL) formula. If uncontrollable propositions $\mu^{\text{uc}}$ are excluded, i.e., $M^{\text{uc}}=\emptyset$, then  $\phi$ is called a risk signal interval temporal logic (RiSITL) formula. If all risk predicates are replaced by deterministic predicates and $M^{\text{uc}}=\emptyset$, then  $\phi$ reduces to an SITL formula as in \cite{maler2004monitoring}. 

\begin{center}
	\begin{tabular}{ |l|l| } 
		\hline
		\footnotesize{Abbreviation} & \footnotesize{Features}  \\
		\hline
		\footnotesize{ReRiSITL} & \footnotesize{Predicates $M^{\text{Ri}}$, Uncontrollable Propositions $M^{\text{uc}}$} \\ 
		\footnotesize{RiSITL} &  \footnotesize{Predicates $M^{\text{Ri}}$}  \\ 
		\footnotesize{ReSITL} & \footnotesize{Uncontrollable Propositions $M^{\text{uc}}$}  \\
		\footnotesize{SITL} & \footnotesize{Deterministic Predicates $M^{\text{det}}$ only} \\
		\hline
	\end{tabular}
\end{center}

\subsection{From MITL to Timed Signal Transducer}
\label{sec:mitl_to_ta}

We next define metric interval temporal logic (MITL) \cite{alur1996benefits} which has the advantage that it can be translated into a \emph{timed signal transducer} \cite{ferrere2019real}. We later interpret ReRiSITL formulas as MITL formulas and make use of this translation. Instead of predicates and uncontrollable propositions, MITL considers (controllable) propositions $p\in AP$  where $AP$ is a set of atomic  propositions.  The MITL syntax is hence
\begin{align}\label{eq:full_MITL}
\varphi \; ::= \; \top \; | \; p \; | \;  \neg \varphi \; | \; \varphi' \wedge \varphi'' \; | \; \varphi'  U_I \varphi''\; | \; \varphi'  \underline{U}_I \varphi'' \;
\end{align}
where $\varphi'$ and $\varphi''$ are MITL formulas. Let 
\begin{align*}
\boldsymbol{d}:\mathbb{R}_{\ge 0}\to \mathbb{B}^{|AP|}
\end{align*} 
be a Boolean signal corresponding to truth values of $p\in AP$ over time. Define again the projection of $\boldsymbol{d}$ onto  $p\in AP$ as proj$_p(\boldsymbol{d}):\mathbb{R}_{\ge 0}\to \mathbb{B}$ and let $(\boldsymbol{d},t)\models \varphi$ be the satisfaction relation. The continuous-time semantics of an MITL formula \cite[Sec. 4]{ferrere2019real} are defined as $(\boldsymbol{d},t)\models p$ iff proj$_p(\boldsymbol{d})(t)=\top$ while the other operators are as in Definition \ref{def:ReRiSITL}. An MITL formula $\varphi$ is satisfiable if $\exists \boldsymbol{d}\in \mathcal{F}(\mathbb{R}_{\ge 0},\mathbb{B}^{|AP|})$ such that $(\boldsymbol{d},0)\models\varphi$. Note that the symbols $\varphi$ and $\phi$ are used to distinguish between MITL and ReRiSITL formulas, respectively. 

The translation of $\varphi$ into a timed signal transducer is summarized next and follows \cite{ferrere2019real}.  Let 
\begin{align*}
\boldsymbol{c}:=\begin{bmatrix} c_1 &\hdots &c_O \end{bmatrix}^T\in\mathbb{R}_{\ge 0}^O
\end{align*}
 be a vector of $O$ clock variables that obey the continuous dynamics $\dot{c}_o(t):=1$ with $c_o(0):=0$ for $o\in\{1,\hdots,O\}$. Discrete dynamics occur at instantaneous times in form of clock resets. Let 
 \begin{align*}
r:\mathbb{R}_{\ge 0}^O\to \mathbb{R}_{\ge 0}^O
 \end{align*}
  be a reset function such that $r(\boldsymbol{c})=\boldsymbol{c}'$ where either $c_o'=c_o$ or $c_o'=0$. With a slight abuse of notation, we use $r(c_o)=c_o$ and $r(c_0)=0$. Clocks evolve with time when visiting a state of a timed signal transducer, while clocks may be reset during transitions between states. We define clock constraints as Boolean combinations of conditions of the form $c_o\le k$ and $c_o\ge k$ for some $k\in\mathbb{Q}_{\ge 0}$. Let $\Phi(\boldsymbol{c})$ denote the set of all clock constraints over clock variables in $\boldsymbol{c}$. 
\begin{definition}[Timed Signal Transducer \cite{ferrere2019real}]
	A timed signal transducer is a tuple 
	\begin{align*}
	TST:=(S,s_0,\Lambda,\Gamma,\boldsymbol{c},\iota,\Delta,\lambda,\gamma, \mathcal{A})
	\end{align*} 
	where $S$ is a finite set of states, $s_0$ is the initial state with $s_0\cap S=\emptyset$, $\Lambda$ and $\Gamma$ are a finite sets of input and output variables, respectively, $\iota:S\to\Phi(\boldsymbol{c})$ assigns clock constraints over $\boldsymbol{c}$ to each state, $\Delta$ is a transition relation so that $\delta=(s,g,r,s')\in\Delta$ indicates a transition from $s\in S\cup s_0$ to $s'\in S$ satisfying the guard constraint $g\subseteq \Phi(\boldsymbol{c})$ and resetting the clocks according to $r$; $\lambda:S\cup\Delta\to BC(\Lambda)$ and $\gamma:S\cup\Delta\to BC(\Gamma)$ are input and output labeling functions where $BC(\Lambda)$ and $BC(\Gamma)$ denote the sets of all Boolean combinations over $\Lambda$ and $\Gamma$, respectively, and $\mathcal{A}\subseteq 2^{S\cup \Delta}$ is a generalized B\"uchi acceptance condition.
\end{definition}

A \emph{run} of a $TST$ over an input signal $\boldsymbol{d}:\mathbb{R}_{\ge 0}\to \mathbb{B}^{|\Lambda|}$ is an alternation of time and discrete steps resulting in an output signal $\boldsymbol{y}:\mathbb{R}_{\ge 0}\to \mathbb{B}^{|\Gamma|}$. A time step of duration $\tau\in\mathbb{R}_{>0}$ is denoted by 
\begin{align*}
(s,\boldsymbol{c}(t))\xrightarrow{\tau}(s,\boldsymbol{c}(t)+\tau)
\end{align*}
 with $\boldsymbol{d}(t+t')\models\lambda(s)$, $\boldsymbol{y}(t+t')\models\gamma(s)$, and $\boldsymbol{c}(t+t')\models \iota(s)$ for each $t'\in (0,\tau)$. A discrete step at time $t$ is denoted by 
 \begin{align*}
(s,\boldsymbol{c}(t))\xrightarrow{\delta}(s',r(\boldsymbol{c}(t))))
 \end{align*} 
 for some transition $\delta=(s,g,r,s')\in\Delta$ such that $\boldsymbol{d}(t)\models\lambda(\delta)$, $\boldsymbol{y}(t)\models\gamma(\delta)$, and $\boldsymbol{c}(t)\models g$. Each run starts with a discrete step from the initial configuration $(s_0,\boldsymbol{c}(0))$. Formally, a run of a $TST$ over $\boldsymbol{d}$ is a sequence 
 \begin{align*}
(s_0,\boldsymbol{c}(0))\xrightarrow{\delta_0}(s_1,r_0(\boldsymbol{c}(0)))\xrightarrow{\tau_1}(s_1,r_0(\boldsymbol{c}(0))+\tau_1)\xrightarrow{\delta_1} \hdots.
 \end{align*} 
 Due to the alternation of time and discrete steps, the signals $\boldsymbol{d}(t)$ and $\boldsymbol{y}(t)$ may be a concatenation of sequences consisting of points and open intervals. We associate a function $q:\mathbb{R}_{\ge 0}\to S\cup \Delta$ with a run as $q(0):=\delta_0$, $q(t)=s_1$ for all $t\in(0,\tau_1)$, $\hdots$; $\mathcal{A}$ is a generalized B\"uchi acceptance condition so that a run over $\boldsymbol{d}(t)$ is \emph{accepting} if, for each $A\in\mathcal{A}$, $\text{inf}(q)\cap A\neq \emptyset$ where $\text{inf}(q)$ contains the states in $S$ that are visited, in $q$, for an unbounded time duration and transitions in $\Delta$  that are taken, in $q$, infinitely many times. The language of $TST$ is 
 \begin{align*}
L(TST):=\{\boldsymbol{d}\in\mathcal{F}(\mathbb{R}_{\ge 0}, \mathbb{R}^{|\Lambda|})\big|TST\text{ has an accepting run over } \boldsymbol{d}(t)\}
 \end{align*}  
 The synchronous behavior of two timed signal transducers $TST_1$ and $TST_2$ is defined by their \emph{synchronous product}  $TST_1||TST_2$.  The input-output behavior of $TST_1$ being the input of $TST_2$ is denoted by their \emph{input-output composition} $TST_1\rhd TST_2$, see \cite{ferrere2019real} and \cite[Def. 2 and 3]{lindemann2019efficient} for definitions.

\begin{figure*}[tbh]
	\centering
	\begin{subfigure}{0.32\textwidth}
		\centering
		\includegraphics[scale=0.245]{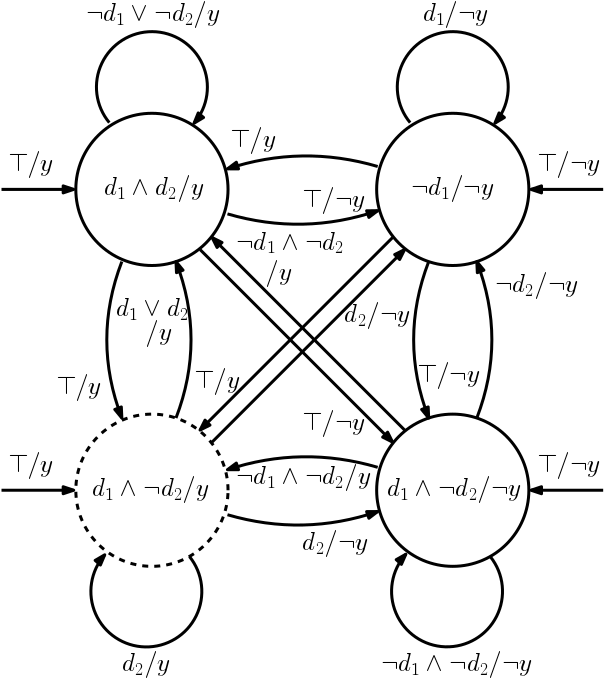}\caption{Timed signal transducer for $U_{(0,\infty)}$}\label{fig:until}
	\end{subfigure}
	\begin{subfigure}{0.32\textwidth}
		\centering
		\includegraphics[scale=0.245]{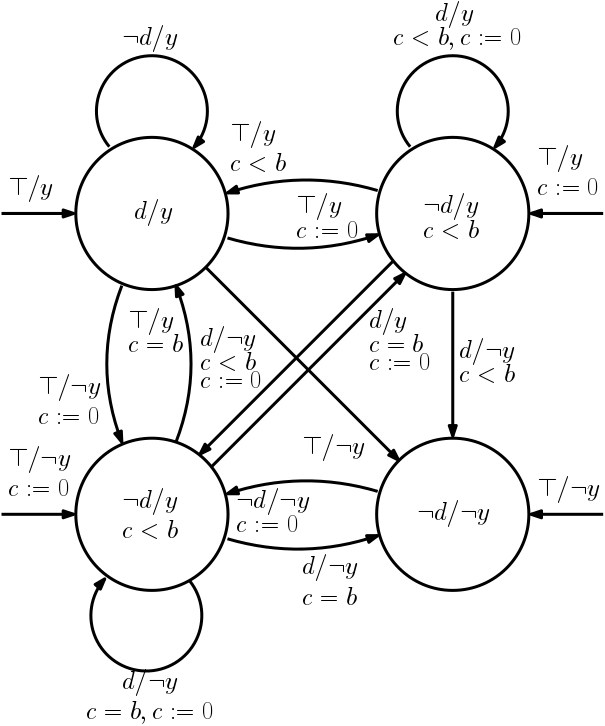}\caption{Timed signal transducer for $F_{(0,b)}$}\label{fig:eventually}
	\end{subfigure}
	\begin{subfigure}{0.32\textwidth}
		\centering
		\includegraphics[scale=0.245]{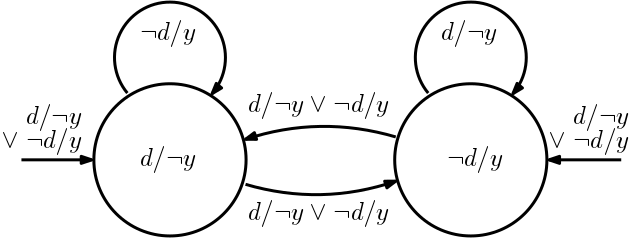}\caption{Timed signal transducer for $\neg$}\label{fig:neg}
	\end{subfigure}\\
	\begin{subfigure}{0.38\textwidth}
		\centering
		\includegraphics[scale=0.245]{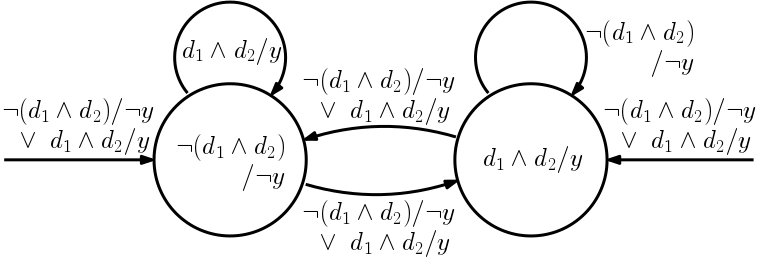}\caption{Timed signal transducer for $\wedge$}\label{fig:conjunction}
	\end{subfigure}
	\begin{subfigure}{0.58\textwidth}
		\centering
		\includegraphics[scale=0.255]{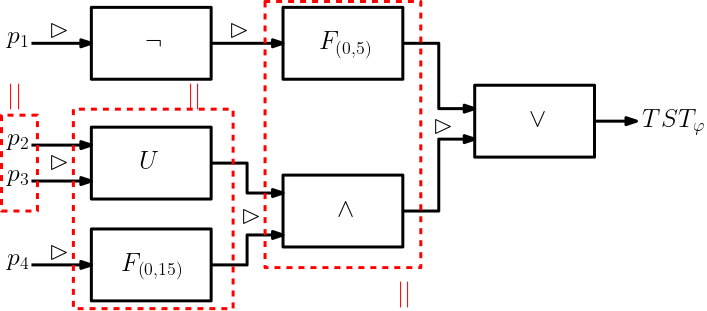}\caption{Formula tree for  $\varphi:=F_{(0,5)} \neg p_1 \vee (p_2 U_{(0,\infty)} p_3 \wedge F_{(0,15)}p_4)$.}\label{fig:formula_tree}
	\end{subfigure}
	\caption{Figs. \ref{fig:until}-\ref{fig:conjunction} show  timed signal transducers for the basic temporal operators $U_{(0,\infty)}$ and $F_{(0,b)}$ and the Boolean operators $\neg$ and $\wedge$. Note that the variables $d$, $d_1$, and $d_2$ here  are used  as generic input symbols, while $y$ is a generic  output symbol. Fig. \ref{fig:formula_tree} shows the formula tree for the MITL formula $\varphi:=F_{(0,5)}\neg p_1\vee (p_2 U_{(0,\infty)}p_3\wedge F_{(0,15)}p_4)$. To construct the timed signal transducer $TST_\varphi$ for $\varphi$ from the formula tree, the synchronous product operation  $||$  and the input-output composition operation $\rhd$ need to be applied to the basic timed signal transducers of the blocks in the formula tree as indicated in Fig. \ref{fig:formula_tree}.  }
	\label{fig:tsd_temporal}
\end{figure*}

We can now summarize the procedure of \cite{ferrere2019real}. First, it is shown that every MITL formula $\varphi$ can be rewritten using only the temporal operators $U_{(0,\infty)}$, $\underline{U}_{(0,\infty)}$,  $F_{(0,b)}$, and $\underline{F}_{(0,b)}$ for rational constants $b$ \cite[Proposition 4.5]{ferrere2019real} using the rewriting rules in \cite[Lemmas 4.1, 4.2, 4.3, and 4.4]{ferrere2019real}. Second, timed signal transducers for $U_{(0,\infty)}$, $\underline{U}_{(0,\infty)}$,  $F_{(0,b)}$, and $\underline{F}_{(0,b)}$ are proposed, see Figs. \ref{fig:until} and \ref{fig:eventually} for examples of $\underline{U}_{(0,\infty)}$ and $F_{(0,b)}$. Note that all states and transitions except for the state indicated by the dashed circle in $U_{(0,\infty)}$ are included in  $\mathcal{A}$. Timed signal transducers for negations and conjunctions are shown in Figs. \ref{fig:neg} and \ref{fig:conjunction}. Third, the formula tree of an MITL formula $\varphi$ is constructed as illustrated in Fig. \ref{fig:formula_tree}.  Fourth, input-output composition $\rhd$ and the synchronous product $||$ are used to obtain a timed signal transducer 
\begin{align*}
TST_\varphi:=(S,s_0,\Lambda,\Gamma,\boldsymbol{c},\iota,\Delta,\lambda,\gamma, \mathcal{A})
\end{align*}
with $\Lambda:=AP$ and $\Gamma:=\{y\}$; $TST_\varphi$ has accepting runs over $\boldsymbol{d}$, i.e., $\boldsymbol{d}\in L(TST_\varphi)$, with $\boldsymbol{y}(0)=\top$ if and only if $(\boldsymbol{d},0)\models\varphi$ \cite[Thm. 6.7]{ferrere2019real}. Note that $\boldsymbol{y}(0)=\top$ (meaning that $\gamma(\delta_0)=y$ where $\delta_0$ is the initial transition) indicates satisfaction of $\varphi$ at time $t=0$, while $\boldsymbol{y}(0)=\bot$, i.e., $\gamma(\delta_0)=\neg y$, indicates $(\boldsymbol{d},0)\not\models\varphi$.

\subsection{Problem Definition}

The first problem is a verification problem to check the satisfiability of an ReRiSITL formula $\phi$ according to Definition~\ref{def:satisfiability}.   
 \begin{problem}\label{prob:1}
	Given a random variable $\boldsymbol{X}$ and an ReRiSITL formula $\phi$ as in \eqref{eq:full_ReRiSTL},  check whether or not $\phi$ is satisfiable. 
\end{problem}

The second problem  is a control problem. Let the system
\begin{align}\label{system_noise}
\dot{\boldsymbol{x}}(t)&=f(\boldsymbol{x}(t))+g(\boldsymbol{x}(t))\boldsymbol{u}, \; \boldsymbol{x}(0):=\boldsymbol{x}_0
\end{align} 
where $f:\mathbb{R}^n\to\mathbb{R}^n$ and $g:\mathbb{R}^n\to\mathbb{R}^{n\times m}$ are locally Lipschitz continuous and where $\boldsymbol{u}\in\mathbb{R}^m$ is a control law. 

In this context, $\boldsymbol{X}$ and $M^{\text{uc}}$  may model  the environment in which the system in \eqref{system_noise} operates, e.g., regions of interest and sensor failures can be modeled by $\boldsymbol{X}$ and  $M^{\text{uc}}$, respectively. Let now each $\mu_m \in M^{\text{Ri}}$ with  $m\in\{1,\hdots,|M^{\text{Ri}}|\}$ be associated with predicate functions $h_m:\mathbb{R}^n\times \mathbb{R}^{\tilde{n}}\to\mathbb{R}$ and risk parameters $R_m(\cdot)$, $\beta_m$, and  $\gamma_m$.  For $\mu^{\text{uc}}\in M^{\text{uc}}$, let  the truth value of $\mu^{\text{uc}}$ at time $t\in\mathbb{R}_{\ge 0}$ be captured by $\boldsymbol{s}\in\mathcal{F}(\mathbb{R}_{\ge 0},\mathbb{B}^{|M^{\text{uc}}|})$, i.e.,  we observe proj$_{\mu^\text{uc}}(\boldsymbol{s})(t)$. Since $\boldsymbol{s}$ is not known beforehand, we assume to observe $\boldsymbol{s}(t)$ at time $t$.

 \begin{problem}\label{prob:2}
Given a random variable $\boldsymbol{X}$ and a satisfiable ReRiSITL formula $\phi$ as in \eqref{eq:full_ReRiSTL},  find a nonanticipative strategy $\boldsymbol{u}(\boldsymbol{x}(t),\boldsymbol{s},t)$ s.t. $(\boldsymbol{x},\boldsymbol{s},\boldsymbol{X},0)\models \phi$ where $\boldsymbol{x}$ is the solution to~\eqref{system_noise} under $\boldsymbol{u}(\boldsymbol{x}(t),\boldsymbol{s},t)$ and where $\boldsymbol{s}(t)$ is observed at time~$t$. 
\end{problem}

The next assumption is not explicitly used and needed for our proposed solutions to Problems \ref{prob:1} and \ref{prob:2}. We will, however, refer to this assumption in some places to emphasize that computational advantages can be obtained under it.

 	\begin{assumption}\label{ass:1}
 		The functions $h_m:\mathbb{R}^n\times \mathbb{R}^{\tilde{n}}\to\mathbb{R}$  are linear in its first argument.
 	\end{assumption}


\section{Satisfiability of ReRiSITL Specifications}
\label{sec:strategy}

In this section, we present a solution to Problem \ref{prob:1}.  In Sections \ref{sec:to_mitl} and \ref{sec:tst_to_reg}, we  construct a timed signal transducer $TST_{\phi}$ that characterizes all signals $\boldsymbol{x}:\mathbb{R}_{\ge 0}\to\mathbb{R}^n$ and $\boldsymbol{s}:\mathbb{R}_{\ge 0}\to\mathbb{B}^{|M^{\text{uc}}|}$ such that $(\boldsymbol{x},\boldsymbol{s},\boldsymbol{X},0)\models \phi$. In Section \ref{sec:satisfiability}, we consider if, for all $\boldsymbol{s}\in\mathcal{F}(\mathbb{R}_{\ge 0},\mathbb{B}^{|M^{\text{uc}}|})$, there exists a nonanticipative strategy $\boldsymbol{x}_\text{na}:\mathcal{F}(\mathbb{R}_{\ge 0},\mathbb{B}^{|M^{\text{uc}}|})\to\mathcal{F}(\mathbb{R}_{\ge 0},\mathbb{R}^n)$ such that $(\boldsymbol{x}_\text{na}(\boldsymbol{s}),\boldsymbol{s},\boldsymbol{X},0)\models \phi$, solving Problem~\ref{prob:1}. 

\subsection{From ReRiSITL to Timed Signal Transducer}
\label{sec:to_mitl}
The first goal is  to abstract the ReRiSITL formula $\phi$ into an MITL formula $\varphi$ via a transformation $Tr(\cdot)$.  Therefore, let us use the notation $\phi(M)$  to make explicit that the  ReRiSITL formula $\phi$ depends on the set of predicates and  propositions  $M$. The transformation $Tr(\cdot)$  essentially replaces predicates and uncontrollable propositions $M$ in $\phi(M)$ by a set of propositions $AP$.  For $i\in\{1,\hdots,|M|\}$, associate with each $\mu_i\in M$  a proposition $p_i$ and let $AP:=\{p_1,\hdots,p_{|M|}\}$. Let then 
\begin{align*}
\varphi:=Tr(\phi(M))=\phi(AP),
\end{align*}
 e.g., $\phi(M):=F_{I}(\mu_1 \wedge \mu_2)$ becomes $\varphi:=\phi(AP)=F_{I}(p_1\wedge p_2)$. Let the inverse 
 \begin{align*}
Tr^{-1}(\varphi)=Tr^{-1}(Tr(\phi(M)))=\phi(M)
 \end{align*}
  be obtained by replacing each $p_i\in AP$ in $\varphi$ with the corresponding $\mu_i\in M$.

Let now  $TST_\varphi:=(S,s_{0},\Lambda,\Gamma,\boldsymbol{c},\iota,\Delta,\lambda,\gamma, \mathcal{A})$ be constructed for the MITL formula $\varphi$ according to Section \ref{sec:mitl_to_ta} with $\Lambda:=AP$. Since we aim at satisfying the STL formula $\phi$, we modify $TST_\varphi$ by the following operations to account for the error induced by the abstraction from $\phi$ to $\varphi$ via $Tr$.
\begin{enumerate}
	\item[{[O1]}]  Remove each state $s\in S$ for which there exists no $\boldsymbol{x}\in\mathbb{R}^n$ and no $\boldsymbol{s}\in\mathbb{B}^{|M^{\text{uc}}|}$ so that $(\boldsymbol{x},\boldsymbol{s},\boldsymbol{X})\models Tr^{-1}(\lambda(s))$.\footnote{We use $(\boldsymbol{x},\boldsymbol{s},\boldsymbol{X})\models Tr^{-1}(\lambda(s))$ with a slight abuse of notation  instead of $(\boldsymbol{x},\boldsymbol{s},\boldsymbol{X},t)\models Tr^{-1}(\lambda(s))$ since $Tr^{-1}(\lambda(s))$ is  a Boolean formula.} Remove the corresponding $s$ from $\mathcal{A}$. Further remove the corresponding ingoing ($(s',g,r,s)\in\Delta$ for some $s'\in S$) and outgoing ($(s,g,r,s')\in\Delta$ for some $s'\in S$) transitions.
	\item[{[O2]}]  Remove each transition $\delta:=(s,g,r,s')\in\Delta$ for which there exists no $\boldsymbol{x}\in\mathbb{R}^n$ and no $\boldsymbol{s}\in\mathbb{B}^{|M^{\text{uc}}|}$  so that $(\boldsymbol{x},\boldsymbol{s},\boldsymbol{X})\models Tr^{-1}(\lambda(\delta))$. Remove the corresponding $\delta$ from $\mathcal{A}$.
\end{enumerate}

The modified $TST_\varphi$ is denoted by 
\begin{align*}
TST_{\phi}:=(S^{\phi},s_{0},\Lambda,\Gamma,\boldsymbol{c},\iota,\Delta^{\phi},\lambda,\gamma, \mathcal{A}^{\phi})
\end{align*}
 for which naturally $S^{\phi}\subseteq S$, $\Delta^{\phi}\subseteq \Delta$, and $\mathcal{A}^{\phi}\subseteq \mathcal{A}$. Note that it is essential to be able to check  if there exists $\boldsymbol{x}\in\mathbb{R}^n$ and $\boldsymbol{s}\in\mathbb{B}^{|M^{\text{uc}}|}$ such that $(\boldsymbol{x},\boldsymbol{s},\boldsymbol{X})\models Tr^{-1}(\lambda(s))$ and $(\boldsymbol{x},\boldsymbol{s},\boldsymbol{X})\models Tr^{-1}(\lambda(\delta))$ in [O1] and [O2], respectively. To do so, techniques  as  in \cite{williams2013model} and summarized in \cite[Ch. 2]{bemporad1999control}, resulting in nonlinear mixed integer programs, can be employed. Nonlinearity here is in particular induced due to  $R(\cdot)$. To address Problem \ref{prob:2} (which will also rely on operations [O1] and [O2]), addressed in Sections \ref{sec:determinization} and \ref{sec:planning}, we will obtain computationally more efficient mixed integer linear programs if Assumption \ref{ass:1} holds.   

\subsection{Satisfiability of RiSITL Specifications}
\label{sec:tst_to_reg}
To characterize all signals $\boldsymbol{x}:\mathbb{R}_{\ge 0}\to \mathbb{R}^n$ and $\boldsymbol{s}:\mathbb{R}_{\ge 0}\to\mathbb{B}^{|M^{\text{uc}}|}$ so that $(\boldsymbol{x},\boldsymbol{s},\boldsymbol{X},0)\models \phi$, we translate $TST_{\phi}$ of the previous subsection, which is in essence a timed automaton  when removing the output labels, to a region automaton $RA(TST_{\phi})$ \cite{alur1994theory}\footnote{We could equivalently use the computationally-efficient zone automaton, which is avoided here to keep the discussion in the remainder simple.}; $RA(TST_{\phi})$ can be used to check emptiness of $TST_{\phi}$, i.e., to analyze reachability properties of $TST_{\phi}$. Since $TST_{\phi}$ has invariants on states $\iota(s)$ and guards $g$ included in transitions  $(s,g,r,s')\in\Delta^{\phi}$, we have to slightly modify the algorithms presented in \cite{alur1994theory,alur1996benefits}. Therefore, we associate a transition relation $\Rightarrow$ over the extended state space $S^{\phi}\times \mathbb{R}_{\ge 0}^O$.
\begin{definition}[Equivalent transition system of $TST_{\phi}$]\label{def:trans_rel}
	 Let $(S^{\phi}\times \mathbb{R}_{\ge 0}^O,\Rightarrow)$ be a transition system with $(s,\boldsymbol{c})\xRightarrow{\delta}(s',\boldsymbol{c}')$ if and only if there exist $t'\in\mathbb{R}_{\ge 0}$ and $\delta:=(s,g,r,s')\in\Delta^{\phi}$ so that
	\begin{itemize}
		\item for all $\tau\in(0,t')$, $\boldsymbol{c}+\tau\models\iota(s)$,
		\item it holds that $\boldsymbol{c}'=r(\boldsymbol{c}+t')$ and  $\boldsymbol{c}+t'\models g$,
	\end{itemize}
	i.e., a combination of time and discrete transitions.
\end{definition} 

Reachability properties of the infinite state transition system $(S^{\phi}\times \mathbb{R}_{\ge 0}^O,\Rightarrow)$ (and hence of $TST_{\phi}$) can now be analyzed by its finite state region automaton $RA(TST_{\phi})$ that relies on a bisimulation relation $\sim\subseteq \mathbb{R}^O_{\ge 0}\times \mathbb{R}^O_{\ge 0}$ resulting in clock regions. In fact, a clock region is an equivalence class induced by $\sim$.  Details are omitted and the reader is referred to \cite{alur1994theory} for details on the bisimulation $\sim$ and on clock regions. Let $\alpha$ and $\alpha'$ be clock regions and assume $\boldsymbol{c}\in\alpha$ and $\boldsymbol{c}'\in\alpha'$. If $(s,\boldsymbol{c})\xRightarrow{\delta}(s',\boldsymbol{c}')$ and $\boldsymbol{c}\sim \bar{\boldsymbol{c}}$ for some $\bar{\boldsymbol{c}}$, then it  holds that there is a $\bar{\boldsymbol{c}}'$ with $\boldsymbol{c}'\sim \bar{\boldsymbol{c}}'$ so that  $(s,\bar{\boldsymbol{c}})\xRightarrow{\delta}(s',\bar{\boldsymbol{c}}')$.

\begin{definition}[Region automaton of $TST_{\phi}$]\label{def:region}
	The region automaton 
	\begin{align*}
	RA(TST_{\phi}):=(Q,q_0,\Delta_R,\mathcal{A}_R)
	\end{align*}
	 is the quotient system of $(S^{\phi}\times \mathbb{R}_{\ge 0}^O,\Rightarrow)$ using clock regions as equivalence classes and  defined as:
	\begin{itemize}
		\item The states are $q:=(s,\alpha)$ where $s\in S^{\phi}$ and $\alpha\in A$ where $A$ is the set of all clock regions so that $Q:=S^{\phi}\times A$.
		\item The initial states are $q_0:=(s_0,\alpha_0)\in Q$ where $\alpha_0$ is the clock region corresponding to $\boldsymbol{c}(0)$.
		\item For $q:=(s,\alpha)$ and $q':=(s',\alpha')$, there is a transition $(q,\delta, q')\in\Delta_R$ if and only if there is a transition $(s,\boldsymbol{c})\xRightarrow{\delta}(s',\boldsymbol{c}')$ for $\boldsymbol{c}\in\alpha$ and $\boldsymbol{c}'\in\alpha'$.
		\item $q=(s,\alpha)\in \mathcal{A}_{R}(i)$  if $s\in\mathcal{A}^{\phi}(i)$.
	\end{itemize}
\end{definition}

Using standard graph search techniques such as the memory efficient variant of the nested depth first search \cite{courcoubetis1992memory}, here adapted to deal with the generalized B\"uchi acceptance condition as in \cite{tauriainen2006nested}, we may obtain, if existent, and accepting sequence $\mathfrak{q}=(q_0,q_1,\hdots)$ with $q_j:=(s_j,\alpha_j)$ and $(q_j,\delta_j,q_{j+1})\in \Delta_R$ for each $j\in \mathbb{N}$ satisfying the generalized B\"uchi acceptance condition $\mathcal{A}_R$. In particular, $\mathfrak{q}:=(\mathfrak{q}_p,\mathfrak{q}_s^\omega)$ consists of a prefix of length $p+1$ and a suffix of length $s$, here denoted by $\mathfrak{q}_p:=(q_0, \hdots, q_p)$ and  $\mathfrak{q}_s:=(q_{p+1},\hdots,q_{p+s}) $. Furthermore, we require that $\gamma(\delta_0)=y$ to indicate that we want $(\boldsymbol{d},0)\models \varphi$. We next add timings $\bar{\tau}:=(\bar{\tau}_p,\bar{\tau}_s^\omega)$ to $\mathfrak{q}$ with  $\bar{\tau}_p:=(\tau_0:=0,\hdots,\tau_p)$ and $\bar{\tau}_s:=(\tau_{p+1},\hdots,\tau_{p+s})$ where $\tau_j\in\mathbb{R}_{>0}$  for $j \ge 1$ corresponds to the occurence of $\delta_j$, which happens $\tau_j$ time units after the occurence of $\delta_{j-1}$. We have presented a way to find such $\bar{\tau}$  in \cite[Sec. III.C]{lindemann2020control}.

By denoting $T_j:=\sum_{k=0}^j \tau_j$,  $\mathfrak{q}$ and $\bar{\tau}$ can be associated with a \emph{plan} given by
\begin{align}\label{eq:plan1133}
d_p(t):=\begin{cases}
\lambda(\delta_j) &\text{if } t= T_j\\ 
\lambda(s_j) &\text{if }  T_j < t <T_{j+1}
\end{cases}
\end{align}

The intuition of a plan $d_p:\mathbb{R}_{\ge 0}\to BC(AP)$ is as follows:  a signal $\boldsymbol{d}:\mathbb{R}_{\ge 0}\to\mathbb{B}^{|AP|}$ that satisfies the plan $d_p$ also satisfies the MITL specification $\varphi$ at time $t=0$, i.e.,  $\boldsymbol{d}(t)\models d_p(t)$ for all $t\ge 0$ implies that $(d,0)\models\varphi$.

\begin{lemma}\label{lem:1}
	Given a signal $\boldsymbol{d}:\mathbb{R}_{\ge 0}\to \mathbb{B}^{|AP|}$, there is an accepting run of $TST_{\phi}$ over $\boldsymbol{d}(t)$ and $(\boldsymbol{d},0)\models \varphi$ if only if there exists a plan $d_p(t)$ so that  $\boldsymbol{d}(t) \models d_p(t)$ for all $t\in\mathbb{R}_{\ge 0}$.
	
	\begin{proof}
	$\Rightarrow$: Departing from $TST_{\phi}$, the infinite state transition system $(S^{\phi}\times \mathbb{R}_{\ge 0}^O,\Rightarrow)$ has, by construction, the same reachable set as $TST_{\phi}$, i.e, the same reachable configurations 
	\begin{align*}
	(s_0,\boldsymbol{c}(0)),(s_0,r(\boldsymbol{c}(0))), (s_1,r(\boldsymbol{c}(0))+\tau_1),\hdots.
	\end{align*}
	Since $\sim$ is a bisimulation relation, reachability properties of $TST_{\phi}$ can then equivalently be analyzed by considering the finite state transition system $RA(TST_{\phi})$ \cite[Lemma 4.13]{alur1994theory}. If there hence exists an accepting run of $TST_{\phi}$ over $\boldsymbol{d}(t)$ and $(\boldsymbol{d},0)\models \varphi$, i.e., $\gamma(\delta_0)=y$, the plan $d_p(t)$ can be constructed as described above by obtaining $\mathfrak{q}$ and $\bar{\tau}$ directly from the accepting run of $TST_{\phi}$ over $\boldsymbol{d}(t)$. It will, by construction, hold that $\boldsymbol{d}(t)\models d_p(t)$ for all $t\in\mathbb{R}_{\ge 0}$.
	
	$\Leftarrow$: If there exists a plan $d_p(t)$ so that $\boldsymbol{d}(t)\models d_p(t)$ for all $t\in\mathbb{R}_{\ge 0}$, then it follows that $TST_{\phi}$ has an accepting run over $\boldsymbol{d}(t)$. This follows by construction of $d_p(t)$ where $\mathfrak{q}$ and $\bar{\tau}$ have been obtained based on $RA(TST_{\phi})$ (as described for the synthesis of $d_p(t)$) and by the bisimulation relation $\sim$. Removing states and transitions from $TST_\varphi$ according to operations [O1] and [O2] resulting in $TST_{\phi}$ only removes behavior from $TST_\varphi$ (not adding additional behavior), i.e., $L(TST_{\phi}) \subseteq L(TST_\varphi)$, so that, by \cite[Thm. 6.7]{ferrere2019real}, an accepting run of $TST_{\phi}$ over $\boldsymbol{d}(t)$ inducing $\boldsymbol{y}(0)=\top$ results in $(\boldsymbol{d},0)\models \varphi$. 
	\end{proof}
\end{lemma}

Note that there may exist an accepting run of $TST_\varphi$ over $\boldsymbol{d}(t)$ so that $(\boldsymbol{d},0)\models \varphi$, while there exists no accepting run of $TST_{\phi}$ over $\boldsymbol{d}(t)$ due to operations [O1] and [O2]. We can now associate $d_\mu:\mathbb{R}_{\ge 0}\to BC(M)$ with $d_p(t)$ as
\begin{align*}
d_\mu(t):=Tr^{-1}(d_p(t))
\end{align*}
and, based on ${\phi}$, state under which conditions $d_p(t)$ exists.

\begin{theorem}\label{thm:1}
	There exists a plan $d_p(t)$ (and hence a plan $d_\mu(t)$) if and only if there exists $\boldsymbol{x}:\mathbb{R}_{\ge 0}\to \mathbb{R}^n$ and $\boldsymbol{s}:\mathbb{R}_{\ge 0}\to\mathbb{B}^{|M^{\text{uc}}|}$ so that $(\boldsymbol{x},\boldsymbol{s},\boldsymbol{X},0)\models \phi$.
	
	\begin{proof}
	$\Rightarrow$: The existence of a plan $d_p(t)$ implies, by Lemma \ref{lem:1}, that a signal $\boldsymbol{d}:\mathbb{R}_{\ge 0}\to\mathbb{B}^{|AP|}$ with $\boldsymbol{d}(t)\in d_p(t)$ for all $t\in\mathbb{R}_{\ge 0}$ is such that $(\boldsymbol{d},0)\models \varphi$. Operations [O1] and [O2] remove all states $s$ and transitions $\delta$ from $TST_\varphi$ that are infeasible, i.e., for which there exists no $\boldsymbol{x}\in\mathbb{R}^n$ and no $\boldsymbol{s}\in\mathbb{B}^{|M^{\text{uc}}|}$ such that $(\boldsymbol{x},\boldsymbol{s},\boldsymbol{X})\models Tr^{-1}(\lambda(s))$ and $(\boldsymbol{x},\boldsymbol{s},\boldsymbol{X})\models Tr^{-1}(\lambda(\delta))$, respectively. Recall  that the only difference between the semantics of ${\phi}$ and $\varphi$ is the difference in  $\mu_i$ and $p_i$, respectively. It follows that, based on the run of $TST_\varphi$ over $\boldsymbol{d}(t)$, we  can construct a signal $\boldsymbol{x}:\mathbb{R}_{\ge 0}\to\mathbb{R}^n$ and $\boldsymbol{s}:\mathbb{R}_{\ge 0}\to\mathbb{B}^{|M^{\text{uc}}|}$ with $(\boldsymbol{x}(t),\boldsymbol{s}(t),\boldsymbol{X})\models d_\mu(t)$ for all $t\in\mathbb{R}_{\ge 0}$  implying that $(\boldsymbol{x},\boldsymbol{s},\boldsymbol{X},0)\models {\phi}$.
	
	$\Leftarrow$: Based on $\boldsymbol{x}(t)$ and $\boldsymbol{s}(t)$, define the signal 
	\begin{align*}
	\boldsymbol{d}(t):=\begin{bmatrix}
	h_1^\top(\boldsymbol{x}(t)) & \hdots & h_{|M^{\text{c}}|}^\top(\boldsymbol{x}(t)) & \boldsymbol{s}(t)^T
	\end{bmatrix}^T
	\end{align*}
	 where $h_m^\top(\boldsymbol{x}):=\top$ if $R_m(h_m(\boldsymbol{x},\boldsymbol{X}))\le \gamma_m$  and $h_i^\top(\boldsymbol{x}):=\bot$ otherwise and that is such that $(\boldsymbol{d},0)\models\varphi$. Note that $h_m(\boldsymbol{x},\tilde{\boldsymbol{\mu}})$ is the predicate function associated with $\mu_m$. It follows that $\boldsymbol{d}$ induces an accepting run of $TST_{\phi}$  over $\boldsymbol{d}(t)$ since the traversed states and transitions during this run have not been removed by operations [O1] and [O2]. By Lemma \ref{lem:1}, it follows that there hence exists a plan $d_p(t)$. 
	\end{proof}
\end{theorem}

The next two results are straightforward consequences of the previous result.
\begin{corollary}\label{cor:main}
	If $\boldsymbol{x}:\mathbb{R}_{\ge 0}\to \mathbb{R}^n$ and $\boldsymbol{s}:\mathbb{R}_{\ge 0}\to\mathbb{B}^{|M^{\text{uc}}|}$ are so that $(\boldsymbol{x}(t),\boldsymbol{s}(t),\boldsymbol{X})\models d_\mu(t)$ for all $t\in\mathbb{R}_{\ge 0}$, then it follows that $(\boldsymbol{x},\boldsymbol{s},\boldsymbol{X},0)\models {\phi}$.
\end{corollary}
\begin{corollary}\label{cor:main1}
	If $M^{\text{uc}}=\emptyset$, i.e., $\phi$ is an RiSITL formula, then it holds that there exists a plan $d_p(t)$ (and hence a plan $d_\mu(t)$) if and only if $\phi$ is satisfiable.
\end{corollary}

\subsection{Satisfiability of ReRiSITL Specifications}
\label{sec:satisfiability}

The previous results can only be used to check satisfiability of RiSITL. For ReRiSITL specifications $\phi$, this requires to check all $ \boldsymbol{s}\in\mathcal{F}(\mathbb{R}_{\ge 0},\mathbb{B}^{|M^{\text{uc}}|})$ as in Definition \ref{def:satisfiability}. Let us define  
\begin{align*}
\boldsymbol{s}^\bot:=\begin{bmatrix}\bot & \hdots & \bot\end{bmatrix}^T\in \mathbb{B}^{|M^{\text{uc}}|}
\end{align*} 
and additionally impose the following assumption that all signals $ \boldsymbol{s}\in\mathcal{F}(\mathbb{R}_{\ge 0},\mathbb{B}^{|M^{\text{uc}}|})$ have to satisfy.
\begin{assumption}\label{ass:2}
	Assume that  $\boldsymbol{s}(t)=\boldsymbol{s}^\bot$ for all times except on a set of measure zero, i.e., $\boldsymbol{s}(t)\neq \boldsymbol{s}^\bot$ only for a countable set of times $t$. There exists a known lower bound $\zeta>0$ between events $\boldsymbol{s}(t)\neq \boldsymbol{s}^\bot$, i.e., for  $\boldsymbol{s}(t')=\boldsymbol{s}(t'')\neq\boldsymbol{s}^\bot$ with $t'\neq t''$, it holds that $|t''-t'|\ge \zeta$. 
\end{assumption} 

Assumption \ref{ass:2} excludes signals $\boldsymbol{s}(t)$ exhibiting Zeno behavior, i.e., infinite changes of $\boldsymbol{s}(t)$ in finite time, and is realistic in the sense that it allows to model instantaneous error signals such as considered for communication dropouts or sensor failures. Assumption \ref{ass:2} is in particular necessary for a game-based approach, see \cite{maler1995synthesis}. Furthermore, Assumption~\ref{ass:2} is necessary for the replanning procedure in Section \ref{sec:CEGIS}.

\begin{figure}
	\centering
	\includegraphics[scale=0.4]{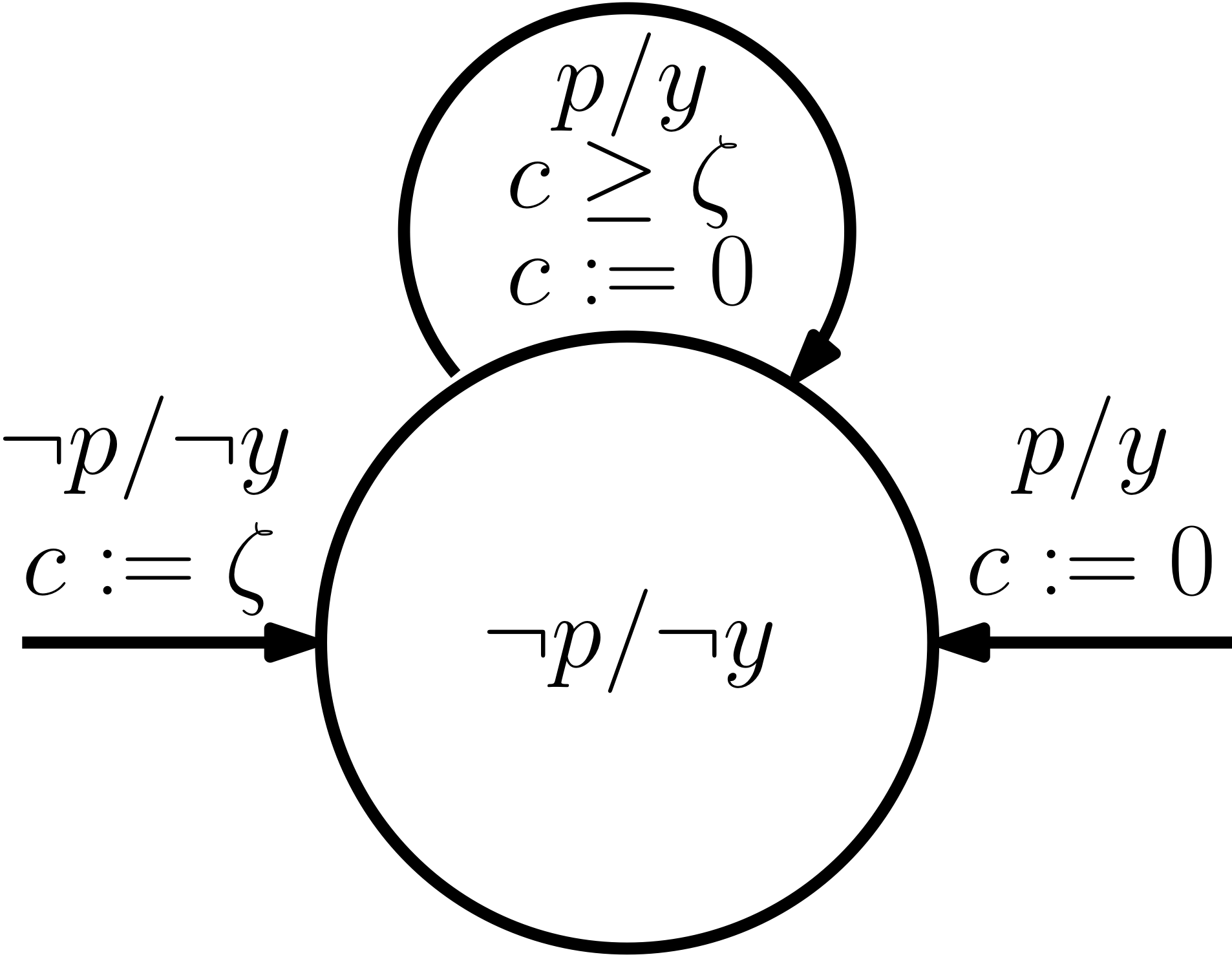}\caption{Timed signal transducer for uncontrollable propositions according to Assumption \ref{ass:2}.}\label{fig:uncontrollable_prop}
\end{figure}

In Algorithm \ref{alg:2}, presented below and explained in the remainder, we summarize the steps to check if $\phi$ is satisfiable. Line 1 in Algorithm \ref{alg:2} has already been explained, while line 2 is related to Assumption \ref{ass:2}. In particular, to model uncontrollable propositions $\mu^\text{uc}\in M^\text{uc}$ according to Assumption \ref{ass:2}, we consider the timed signal transducer in Fig. \ref{fig:uncontrollable_prop}. When constructing $TST_\varphi$, we hence model each $p\in AP$ with $\mu^\text{uc}=Tr^{-1}(p)\in  M^\text{uc}$ as in Fig. \ref{fig:uncontrollable_prop}. Line 3 in Algorithm~\ref{alg:2} then performs [O1] and [O2] to obtain $TST_\phi$.

\begin{algorithm}\label{alg:2}
	\caption{Algorithm to check if $\phi$ is satisfiable.}
	\label{alg:2}
	\begin{algorithmic}[1]
		\State Obtain the MITL formula $\varphi:=Tr(\phi)$.
		\State Obtain $TST_\varphi$ according to Section \ref{sec:mitl_to_ta} and where uncontrollable propositions $p_i\in AP$, i.e., $p_i$ with $Tr^{-1}(p_i)\in M \cap M^\text{uc}$, are modeled as in Fig. \ref{fig:uncontrollable_prop}.
		\State Perform [O1] and [O2] to obtain $TST_\phi$.
		\State Modify $TST_\phi$ to avoid Zeno behavior.
		\State Translate  $TST_\phi$ into ${RA}_C(TST_{\phi})$.
		\State Translate ${RA}_C(TST_{\phi})$ into $\overline{RA}_C(TST_{\phi})$.
		\State Run Algorithm \ref{alg:1} to obtain $W$.
		\State Check if the conditions in Theorem \ref{thm:2} are satisfied.
	\end{algorithmic}
\end{algorithm}

Within the presented game-based approach, it needs to be ensured that no player (here the two players are the controllable and uncontrollable signals $\boldsymbol{x}$ and $\boldsymbol{s}$) wins by inducing Zeno behaviour (see \cite{maler1995synthesis} for more intuition). A generic way of avoiding Zeno behavior is to add a clock $c$ to $TST_\phi$ and add, to each transition, the constraint $c\ge \epsilon$ for a small constant $\epsilon\in\mathbb{Q}_{>0}$ and the reset function $r(c):=0$. This modification will affect the completeness, but not the soundness of the proposed approach. There are minimally invasive algorithms how to avoid Zeno behavior, for instance as in \cite{gomez2007efficient}.  This modification of  $TST_{\phi}$ is stated in line 4 in Algorithm \ref{alg:2}.

	Recall from Section \ref{sec:mitl_to_ta} that an accepting run in $TST_{\phi}$ needs to satisfy the generalized B\"uchi acceptance condition which implies having infinite length, i.e., the run is not allowed to stop existing.  The latter is necessary since we require to be able to extend each finite run in $TST_{\phi}$  to an infinite run. Specifically, note that within a state $s\in S^{\phi}$ in $TST_{\phi}$ it may happen that, for some $\boldsymbol{s}\in \mathbb{B}^{|M^{\text{uc}}|}$, there exists no $\boldsymbol{x}\in\mathbb{R}^n$ such that a transition can be taken, i.e.,  there exists no $\delta':=(s,g',r',s'')\in\Delta^{\phi}$  such that $(\boldsymbol{x},\boldsymbol{s},\boldsymbol{X})\models Tr^{-1}(\lambda(\delta'))$. This means that there is no continuation of a finite run entering the state $s$ so that the run is not accepting. For instance, in Fig. \ref{fig:eventually} in the bottom right state there exists no transition for proj$_d(\boldsymbol{d})=\top$. 
%
%
To account for this, we first modify the infinite state transition system $(S^{\phi}\times \mathbb{R}_{\ge 0}^O,\Rightarrow)$ to $(S^{\phi}\times \mathbb{R}_{\ge 0}^O,\Rightarrow_C)$ by separating time and  discrete transitions. 
\begin{definition}[Equivalent transition system of $TST_{\phi}$]\label{def:trans_rel_}
	Let $(S^{\phi}\times \mathbb{R}_{\ge 0}^O,\Rightarrow_C)$ be a transition system where $(s,\boldsymbol{c})\xRightarrow{\delta_t}_C (s',\boldsymbol{c}')$ with $\delta_t\in\{\delta,t\}$ if  there is either a \emph{discrete or a time transition} as follows:
	\begin{enumerate}
		\item there is a \emph{discrete transition}  $(s,\boldsymbol{c})\xRightarrow{\delta}_C (s',\boldsymbol{c}')$  if there exists $\delta:=(s,g,r,s')\in\Delta^{\phi}$ so that $\boldsymbol{c}'=r(\boldsymbol{c})$ and  $\boldsymbol{c}\models g$,
		\item there is a \emph{time transition}  $(s,\boldsymbol{c})\xRightarrow{t}_C (s,\boldsymbol{c}')$ if, for all $\tau\in(0,t)$, $\boldsymbol{c}+\tau\models\iota(s)$.
	\end{enumerate}
\end{definition}


We emphasize that $(S^{\phi}\times \mathbb{R}_{\ge 0}^O,\Rightarrow_C)$, $(S^{\phi}\times \mathbb{R}_{\ge 0}^O,\Rightarrow)$, and hence $TST_\phi$ have the same reachability properties. Let now $RA_C(TST_{\phi}):=(Q,q_0,\Delta_R,\mathcal{A}_R)$ denote the region automaton, similar to Definition \ref{def:region}, but now obtained from  $(S^{\phi}\times \mathbb{R}_{\ge 0}^O,\Rightarrow_C)$ instead of $(S^{\phi}\times \mathbb{R}_{\ge 0}^O,\Rightarrow)$. The translation to $RA_C(TST_{\phi})$ corresponds to line 5 in Algorithm~\ref{alg:2}.
\begin{definition}[Region automaton of $TST_{\phi}$]\label{def:region_}
	The region automaton 
	\begin{align*}
	RA_C(TST_{\phi}):=(Q,q_0,\Delta_R,\mathcal{A}_R)
	\end{align*} 
	is   defined as:
	\begin{itemize}
		\item The states are $q:=(s,\alpha)$ where $s\in S^{\phi}$ and $\alpha\in A$ where $A$ is the set of all clock regions so that $Q:=S^{\phi}\times A$.
		\item The initial states are $q_0:=(s_0,\alpha_0)\in Q$ where $\alpha_0$ is the clock region corresponding to $\boldsymbol{c}(0)$.
		\item For $q:=(s,\alpha)$ and $q':=(s',\alpha')$, there is a transition $(q,\delta_t, q')\in\Delta_R$ where $\delta_t\in\{\delta,t\}$  if  there is
		\begin{enumerate}
			\item either a discrete transition $(s,\boldsymbol{c})\xRightarrow{\delta}_C (s',\boldsymbol{c}')$ for $\boldsymbol{c}\in\alpha$ and $\boldsymbol{c}'\in\alpha'$.
		\item or a time transition $(s,\boldsymbol{c})\xRightarrow{t}_C (s',\boldsymbol{c}')$ for $\boldsymbol{c}\in\alpha$ and $\boldsymbol{c}'\in\alpha'$ where $\alpha'$ is the immediate time successor of $\alpha$\footnote{See \cite[Def. 4.6]{alur1994theory} for the definition of a time successor. By an ``immediate" time successor, we mean that the regions $\alpha$ and $\alpha'$ are connected.}.
		\end{enumerate} 
		\item $q=(s,\alpha)\in \mathcal{A}_{R}(i)$  if $s\in\mathcal{A}^{\phi}(i)$.
	\end{itemize}
\end{definition}

\begin{remark}
	Defining $RA_C(TST_{\phi})$ based on $(S^{\phi}\times \mathbb{R}_{\ge 0}^O,\Rightarrow_C)$ by separating discrete and time transitions, and  unrolling the time domain as in Definition \ref{def:region_}, results in more states compared to  $RA(TST_{\phi})$ based on $(S^{\phi}\times \mathbb{R}_{\ge 0}^O,\Rightarrow)$. This, however, now becomes necessary since uncontrollable signals $\boldsymbol{s}$ may cause undesireable behavior at all times. 
\end{remark}

%
%

To simplify the search of an accepting run in $TST_{\phi}$ via $RA_C(TST_{\phi})$, translate now $RA_C(TST_{\phi})$, which is a finite automaton with \emph{generalized} B\"uchi acceptance condition, into an equivalent finite automaton 
\begin{align*}
\overline{RA}_C(TST_{\phi}):=(\overline{Q},\overline{q}_0,\overline{\Delta}_R,\overline{\mathcal{A}}_R)
\end{align*}
 with a B\"uchi acceptance condition instead, as follows:
\begin{itemize}
	\item $\overline{Q}:=Q\times \{1,\hdots,|\mathcal{A}_R|\}$
	\item $\overline{q}_0:=(q_0,1)$
	\item $\overline{\Delta}_R:=\{((q,i),\delta_t,(q',j))| (q,\delta_t,q')\in\Delta_R \text{ and if } q\in \mathcal{A}_R(i), \text{ then } j=((i+1)\mod |\mathcal{A}_R|+1) \text{ else }j=i   \}$ where $\mathcal{A}_R(i)$ denotes the $i$th element of  $\mathcal{A}_R$
	\item $\overline{\mathcal{A}}_R:=(\mathcal{A}_R(1),1)$.
\end{itemize}
In particular, the difference is that ${\mathcal{A}}_R$ consists of several sets ${\mathcal{A}}_R(i)$ of states, while $\overline{\mathcal{A}}_R$ is a single set of states. By construction, the accepting behavior of $RA_C(TST_{\phi})$ and $\overline{RA}_C(TST_{\phi})$ are the same. This translation corresponds to line 6 in Algorithm~\ref{alg:2} and is performed to obtain a simpler acceptance condition that can be expressed as a fixed point expression as we will see below.  In fact, a winning condition (for a game played between $\boldsymbol{s}$ and $\boldsymbol{x}$) is that always eventually $\bar{\mathcal{A}}_R$ can be visited by each finite run of $\overline{RA}_C(TST_{\phi})$.
\begin{remark}\label{rem:5555}
	The translation to $\overline{RA}_C(TST_{\phi})$ may induce $|Q|\cdot|\mathcal{A}_R|$ states. One can avoid such a state explosion by neglecting the acceptance condition $\mathcal{A}$ for all timed signal transducers in Fig. \ref{fig:tsd_temporal} except for the until operator in Fig. \ref{fig:until} where a B\"uchi acceptance condition is needed.
\end{remark}

In the remainder, we are inspired by the work in \cite{maler1995synthesis}. We first introduce the main operator, the controllable predecessor $\pi:2^{\overline{Q}}\to 2^{\overline{Q}}$. For a certain set $W\subseteq \overline{Q}$, define
\begin{align*}
\pi(W)&:=\{ \overline{q}\in \overline{Q}| \forall \boldsymbol{s}\in \mathbb{B}^{|M^{\text{uc}}|},  \exists  (\overline{q},\delta,\overline{q}')\in\overline{\Delta}_R \text{ s.t. } \text{1) } \overline{q}'\in W, \text{ 2) } \exists \boldsymbol{x}\in\mathbb{R}^n \text{ s.t. } (\boldsymbol{x},\boldsymbol{s},\boldsymbol{X})\models Tr^{-1}(\lambda(\delta)) 
\}
\end{align*}

The intuition is that states in $\pi(W)$ will always allow to enforce a transition into $W$ by a suitable $\boldsymbol{x}$ in one step, no matter of the value of  $\boldsymbol{s}$.  We next present Algorithm \ref{alg:1} to obtain the set $W$ from which we can force to always eventually be within $\overline{\mathcal{A}}_R$. Algorithm \ref{alg:1}, called in line 7 in Algorithm \ref{alg:2},  differs from the algorithm presented in \cite{maler1995synthesis} by the definition of the  controllable predecessor $\pi:2^{\overline{Q}}\to 2^{\overline{Q}}$. 

\begin{algorithm}\label{alg:1}
	\caption{Calculation of the winning set  $W$.}
	\label{alg:1}
		 \hspace*{\algorithmicindent} \textbf{Input: $\overline{RA}_C(TST_{\phi})$ and $\pi:2^{\overline{Q}}\to 2^{\overline{Q}}$}  \\
	\hspace*{\algorithmicindent} \textbf{Output: $W$} 
	\begin{algorithmic}[1]
		\State $W_0:=\overline{Q}$
		\For {$i:=0,1,\hdots$ until $W_{i+1}=W_i$}
		\State $H_0:=\emptyset$
		\For {$j:=0,1,\hdots$ until $H_{j+1}=H_j$}
		\State $H_{j+1}:=\pi(H_j)\cup (\overline{\mathcal{A}}_R \cap \pi(W_i))$
		\EndFor
		\State $W_{i+1}:=H_j$
		\EndFor
		\State $W:=W_i$
	\end{algorithmic}
\end{algorithm}

The algorithm starts with $W_0:=\overline{Q}$ (line 1). For this $W_0$, the inner loop (lines 3-5) calculates all states $H_j$ from which states in $\overline{\mathcal{A}}_R \cap \pi(W_0)$ can be reached, i.e., states in $\overline{\mathcal{A}}_R$ that can be reached and are no deadlock states. For $W_1:=H_j$ (line 6), this inner loop is repeated until eventually obtaining the set of states $W$ that can always eventually be reached.  

The set $W$  tells us if we can let time pass or if a transition according to $\pi(W)$ has to be taken in a particular state.  For $TST_{\phi}$ restricted to $W$ this means that, at no time, an uncontrollable proposition $\boldsymbol{s}$ can  force the system into a state from where the B\"uchi acceptance condition can not be satisfied. The operator $\pi(W)$ then determines which $\boldsymbol{x}$ can be selected in case of a particular $\boldsymbol{s}$.  Note in particular, as similarly analyzed in \cite{maler1995synthesis}, that $W_i$ in Algorithm \ref{alg:1} is monotonically decreasing such that a fixed point, i.e., $W_{i+1}=W_i$, is eventually reached such that Algorithm \ref{alg:1} terminates in a finite number of steps.

\begin{theorem}\label{thm:2}
	If $\boldsymbol{s}$ is according to Assumption \ref{ass:2}, then it holds that the ReRiSITL formula $\phi$ is satisfiable if $\overline{q}_0\in W$ and if there exists $(\overline{q}_0,\delta_0,\overline{q}')\in\overline{\Delta}_R$ with $\gamma(\delta_0)=y$.
	
	\begin{proof}
	 First note that due to the use of the timed signal transducer as in Fig. \ref{fig:uncontrollable_prop}, we account for the form of $\boldsymbol{s}$ as in Assumption \ref{ass:2}. Recall also from Theorem \ref{thm:1} that operations [O1] and [O2] restrict the behavior of $TST_{\phi}$ to the signals $\boldsymbol{x}:\mathbb{R}_{\ge 0}\to \mathbb{R}^n$ and $\boldsymbol{s}:\mathbb{R}_{\ge 0}\to\mathbb{B}^{|M^{\text{uc}}|}$ with $(\boldsymbol{x},\boldsymbol{s},\boldsymbol{X},0)\models \phi$. Note that ${RA}_C(TST_{\phi})$ has, by construction, the same reachable set as $TST_\phi$.  Recall also that $RA_C(TST_{\phi})$ and $\overline{RA}_C(TST_{\phi})$ are equivalent so that reachability properties of $TST_{\phi}$ can equivalently be verified on $\overline{RA}_C(TST_{\phi})$.  We now need to prove that, for each $\boldsymbol{s}\in\mathcal{F}(\mathbb{R}_{\ge 0},\mathbb{B}^{|M^{\text{uc}}|})$ that satisfies Assumption \ref{ass:2}, there is an accepting run in $\overline{RA}_C(TST_{\phi})$ restricted to the states in $W$ that satisfies the B\"uchi acceptance condition. By Algorithm \ref{alg:1}, which is guaranteed to terminate in a  finite number of steps, it is ensured that no state in $W$ is a deadlock and can be continued to another state in $W$. Specifically, it is guaranteed that for each state in $W$ an infinite continuation can be found that satisfies the B\"uchi acceptance condition, no matter how $\boldsymbol{s}(t)$ behaves. Note also that Zeno winning conditions have been excluded by modifying $TST_{\phi}$ to not permit Zeno behavior. Since $\overline{q}_0\in W$ and since there exists $(\overline{q}_0,\delta_0,\overline{q}')\in\overline{\Delta}_R$ with $\gamma(\delta_0)=\top$, it follows that $\phi$ is satisfiable in the sense of Definition \ref{def:satisfiability}.
	 \end{proof}
\end{theorem}   
 
 Note also that Theorem \ref{thm:2} is sufficient. Necessity does not hold due to the modification of $TST_{\phi}$ to avoid Zeno behavior, potentially introducing conservatism.
 
 Finally, we remark that Sections \ref{sec:to_mitl} and \ref{sec:tst_to_reg} use graph search techniques, while Section \ref{sec:satisfiability} follows a game-based approach. One could argue that only the game-based  approach solving Problem~\ref{prob:1} is of interest. We have, however, chosen this particular exposition of our results since we will combine graph search techniques with a game-based approach to address Problem \ref{prob:2} in the following Sections \ref{sec:determinization} and \ref{sec:planning}. 


%
%

\section{From ReRiSITL to ReSITL by Determinizing Risk Predicates}
\label{sec:determinization}

\begin{figure*}
	\centering
	\includegraphics[scale=0.2]{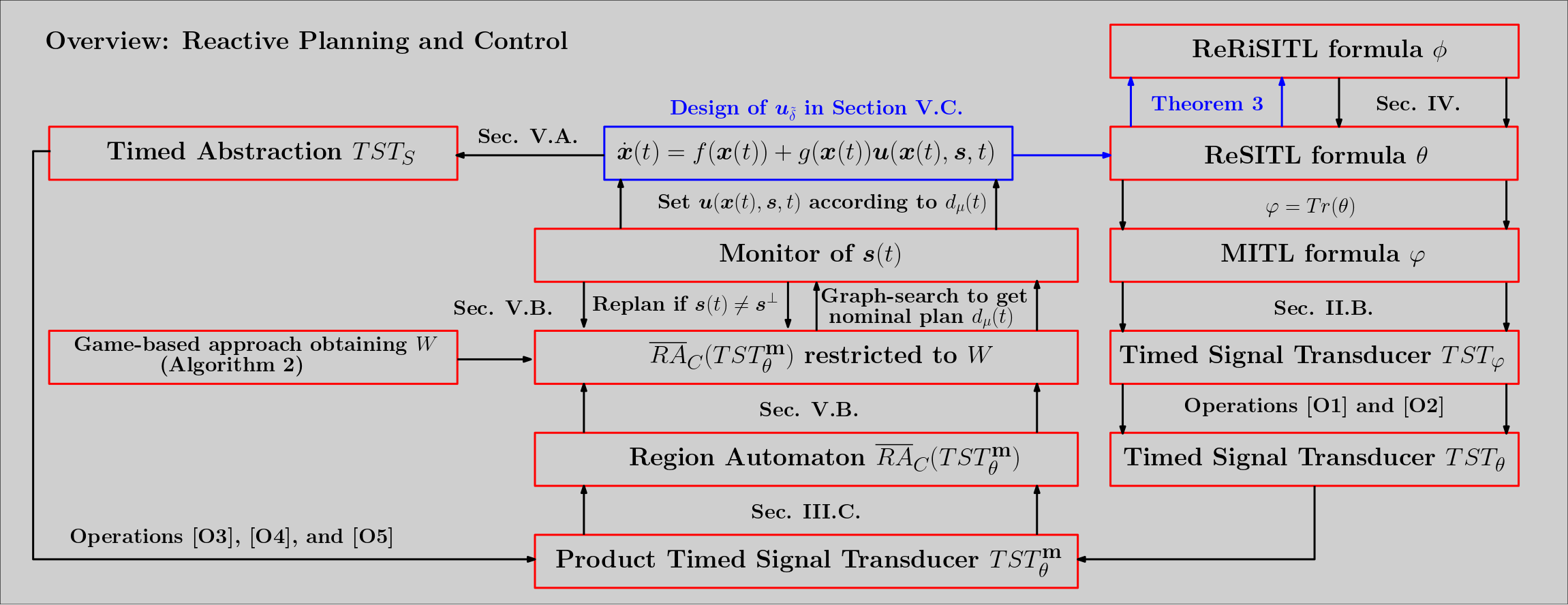}\caption{Overview of the proposed method to reactively plan and control a dynamical system under ReRiSITL Specifications.}\label{fig:overview}
\end{figure*}

Fig. \ref{fig:overview}  can be used as a guide in the remainder as it shows an overview of the reactive planning and control strategy that will be presented in Sections \ref{sec:determinization} and \ref{sec:planning}. Starting in the top right box of Fig. \ref{fig:overview}, this section introduces the idea to determinize risk predicates in $M^{\text{Ri}}$ and replace them with deterministic predicates, hence converting the ReRiSITL formula $\phi$ into an ReSITL formula $\theta$  that we then deal with in Section \ref{sec:planning}. We provide conditions under which a certain soundness property holds which ensures that satisfaction of $\theta$ implies satisfaction of $\phi$. Sections \ref{sec:const_sets} and \ref{sec:determ} assume that $\phi$ is in positive normal form. In the end of Section \ref{sec:determ}, we  discuss how we can deal with siuations where this is not the case.


\subsection{Risk Constrained  Sets}
\label{sec:const_sets}
In the following two sections, we will define risk-tightened  deterministic predicates $\mu_m^\text{det}$ that will replace the risk predicates $\mu_m^\text{ri}$ and allow for the use of existing control methods. Note that  $R(-h_m(\boldsymbol{x},\boldsymbol{X}))$ depends on $\boldsymbol{x}\in\mathbb{R}^n$ (we drop the dependence of $\boldsymbol{x}(t)$ on $t$ in this section for convenience). For given  $\beta_m\in(0,1)$ and $\gamma_m\in\mathbb{R}$,  define the sets
\begin{align*}
\mathfrak{X}_m^{\text{EV}}(\gamma_m)&:=\{\boldsymbol{x}\in\mathfrak{B}|EV[-h_m(\boldsymbol{x},\boldsymbol{X})]\le \gamma_m\},\\
\mathfrak{X}_m^{\text{VaR}}(\beta_m,\gamma_m)&:=\{\boldsymbol{x}\in\mathfrak{B}|VaR_{\beta_m}(-h_m(\boldsymbol{x},\boldsymbol{X}))\le \gamma_m\},\\
\mathfrak{X}_m^{\text{CVaR}}(\beta_m,\gamma_m)&:=\{\boldsymbol{x}\in\mathfrak{B}|CVaR_{\beta_m}(-h_m(\boldsymbol{x},\boldsymbol{X}))\le \gamma_m\}.
\end{align*}
Note the set $\mathfrak{B}\subseteq \mathbb{R}^n$ that is supposed to be an arbitrarily large compact and convex set as will further be explained in Section \ref{sec:control}. The set $\mathfrak{B}$ can be seen as the workspace that \eqref{system_noise} will be forced to remain within. The sets $\mathfrak{X}_m^{\text{EV}}(\gamma_m)$, $\mathfrak{X}_m^{\text{VaR}}(\beta_m,\gamma_m)$, and $\mathfrak{X}_m^{\text{CVaR}}(\beta_m,\gamma_m)$ define all $\boldsymbol{x}$ for which the EV, VaR, and CVaR of $-h_m(\boldsymbol{x},\boldsymbol{X})$  is less or equal than $\gamma_m$, respectively. If these sets are empty, the underlying predicate is not satisfiable. For $c_m\in\mathbb{R}$, which is a design parameter as opposed to  $\beta_m$ and $\gamma_m$, define
\begin{align*}
\mathfrak{X}_m(c_m):=\{\boldsymbol{x}\in\mathfrak{B}|h_m(\boldsymbol{x},\tilde{\boldsymbol{\mu}})\ge c_m\}
\end{align*}
where the mean $\tilde{\boldsymbol{\mu}}$ has been used instead of $\boldsymbol{X}$ to evaluate the predicate function $h_m$.  Note that  $\mathfrak{X}_m(c_m)$ is a compact and convex set if Assumption \ref{ass:1} holds.  If 
\begin{align*}
\mathfrak{X}_m^{\text{EV}}(\gamma_m)&\supseteq \mathfrak{X}_m(c_m), \\
\mathfrak{X}_m^{\text{VaR}}(\gamma_m,\gamma_m)&\supseteq \mathfrak{X}_m(c_m), \\
\mathfrak{X}_m^{\text{CVaR}}(\gamma_m,\gamma_m)&\supseteq \mathfrak{X}_m(c_m),\text{ or }
\end{align*} 
then it holds that 
\begin{align*}
\boldsymbol{x}\in\mathfrak{X}_m(c_m)  &\implies \boldsymbol{x}\in{\mathfrak{X}}_m^{\text{EV}}(\gamma_m), \\
\boldsymbol{x}\in\mathfrak{X}_m(c_m)  &\implies \boldsymbol{x}\in{\mathfrak{X}}_m^{\text{VaR}}(\gamma_m,\gamma_m), \\
\boldsymbol{x}\in\mathfrak{X}_m(c_m)  &\implies \boldsymbol{x}\in{\mathfrak{X}}_m^{\text{CVaR}}(\gamma_m,\gamma_m), \text{ or }\\
\end{align*}
respectively.  This implies that predicates within an ReRiSITL formula $\phi$ can be determinized by using $h_m(\boldsymbol{x},\tilde{\boldsymbol{\mu}})\ge c_m$ (recall \eqref{eq:STL_predicate}) instead of  $R(-h_m(\boldsymbol{x},\boldsymbol{X}))\le \gamma_m$ by conserving an important soundness property (Section \ref{sec:determ}). For given $c_m$, checking these set inclusions may be nonconvex. As shown in \cite[Lemma~1]{lindemann2020control}, when $h_m(\boldsymbol{x},\boldsymbol{X})$ is linear in $\boldsymbol{x}$, this can be checked efficiently since the distribution of $h_m(\boldsymbol{x},\boldsymbol{X})$ is only shifted.
\begin{lemma}{\cite[Lemma~1]{lindemann2020control}}\label{lem:2}
	Assume that $h_m(\boldsymbol{x},\boldsymbol{X})=\boldsymbol{v}^T\boldsymbol{x}+h'(\boldsymbol{X})$ for  $\boldsymbol{v}\in\mathbb{R}^n$ and for  $h':\mathbb{R}^{\tilde{n}}\to\mathbb{R}$, then
	\begin{align*}   
	\mathfrak{X}_m^{\text{EV}}(\gamma_m)\supseteq \mathfrak{X}_m(c_m)&\text{ iff } EV[-h_m(\boldsymbol{x}^*,\boldsymbol{X})]\le \gamma_m \\
	\mathfrak{X}_m^{\text{VaR}}(\beta_m,\gamma_m)\supseteq \mathfrak{X}_m(c_m) &\text{ iff } VaR_{\beta_m}(-h_m(\boldsymbol{x}^*,\boldsymbol{X}))\le \gamma_m \\
	\mathfrak{X}_m^{\text{CVaR}}(\beta_m,\gamma_m)\supseteq \mathfrak{X}_m(c_m) &\text{ iff } CVaR_{\beta_m}(-h_m(\boldsymbol{x}^*,\boldsymbol{X}))\le \gamma_m 
	\end{align*}
	where $\boldsymbol{x}^*:=\underset{\boldsymbol{x}\in\mathfrak{X}_m(c_m)}{\text{argmin}}\; \boldsymbol{v}^T\boldsymbol{x}$  (a convex problem).
\end{lemma}

We remark that in particular $VaR_{\beta_m}(-h_m(\boldsymbol{x}^*,\boldsymbol{X}))$ and $CVaR_{\beta_m}(-h_m(\boldsymbol{x}^*,\boldsymbol{X}))$ can be efficiently computed \cite[Thm. 1]{rockafellar2000optimization} and that $VaR_\beta(-h_m(\boldsymbol{x}^*,\boldsymbol{X}))$ is obtained as a byproduct of the calculation of $CVaR_\beta(-h_m(\boldsymbol{x}^*,\boldsymbol{X}))$. If $h_m(\boldsymbol{x},\boldsymbol{X})$ is nonlinear, we argue that, for some function classes, numerical methods can be used to check these set inclusions, e.g., when $h_m(\boldsymbol{x},\boldsymbol{X})$ is quadratic in $\boldsymbol{x}$.

\subsection{Converting ReRiSITL  into ReSITL Specifications}
\label{sec:determ}

Considering the ReRiSITL formula $\phi$ that consists of the risk predicates $\mu_m\in M^{\text{Ri}}$ with $m\in\{1,\hdots,|M^{\text{Ri}}|\}$, we  transform the ReRiSITL formula  $\phi$ into an ReSITL formula $\theta$. In particular, $\theta$ is obtained by replacing risk predicates $\mu_m\in M^{\text{Ri}}$ in $\phi$ by a deterministic predicate $\mu_m^{\text{det}}$ according to \eqref{eq:STL_predicate}. More formally and by denoting $\phi(M^{\text{Ri}}, M^{\text{uc}})$ instead of $\phi$ to highlight the dependence on risk predicates  $M^{\text{Ri}}$ and uncontrollable propositions $M^{\text{uc}}$, let 
\begin{align*}
\theta:=\phi(M^{\text{det}}, M^{\text{uc}})
\end{align*}
be a ReSITL formula with deterministic predicates
	\begin{align*}
	M^{\text{det}}:=\{\mu_1^{\text{det}},\hdots,\mu_{|M^{\text{Ri}}|}^{\text{det}}\}.
	\end{align*}
Let now 
\begin{align*}
\hat{M}:=M^{\text{det}} \cup M^{\text{uc}}
\end{align*}
be the set of deterministic predicates and uncontrollable propositions. Let us also associate the semantics $(\boldsymbol{x},\boldsymbol{s},\tilde{\boldsymbol{\mu}},t)\models \theta$ with an ReSITL formula $\theta$.\footnote{We define  $(\boldsymbol{x},\boldsymbol{s},\tilde{\boldsymbol{\mu}},t)\models \mu_m^{\text{det}}$ iff $h_m(\boldsymbol{x}(t),\tilde{\boldsymbol{\mu}})\ge c_m$ using \eqref{eq:STL_predicate} instead of  \eqref{eq:risk_predicate}, while the other operators follow as in Section~\ref{sec:ReRiSTL}.} The next assumption is sufficient to ensure soundness in the sense that $(\boldsymbol{x},\boldsymbol{s},\tilde{\boldsymbol{\mu}},t)\models \theta$ implies $(\boldsymbol{x},\boldsymbol{s},\boldsymbol{X},t)\models \phi$.

\begin{assumption}\label{ass:3}
	For each $m\in\{1,\hdots,|M^{\text{Ri}}|\}$,  $\mathfrak{X}_m^{\text{EV}}(\gamma_m)\supseteq \mathfrak{X}_m(c_m)$, $\mathfrak{X}_m^{\text{VaR}}(\beta_m,\gamma_m)\supseteq \mathfrak{X}_m(c_m)$, or $\mathfrak{X}_m^{\text{CVaR}}(\beta_m,\gamma_m)\supseteq \mathfrak{X}_m(c_m)$ (depending on the type of predicate).
\end{assumption}

	\begin{example}\label{ex:3}
		 By setting $c:=0.35$ for the VaR predicates and $c:=0.9$ for the CVaR predicates in Example \ref{ex:2}, Assumption \ref{ass:3} is satisfied. The red circles in Fig. \ref{fig:sim_overview}  indicate the obtained deterministic predicates, based on the predicate functions
		\begin{align*}
		h_\text{R1}^\text{det}(\boldsymbol{x},\tilde{\boldsymbol{\mu}})&:=\epsilon-\|\boldsymbol{x}-\tilde{\boldsymbol{\mu}}_\text{R1}\|^2-0.35\\
		h_\text{R2}^\text{det}(\boldsymbol{x},\tilde{\boldsymbol{\mu}})&:=\epsilon-\|\boldsymbol{x}-\tilde{\boldsymbol{\mu}}_\text{R2}\|^2-0.35\\
		h_\text{O1}^\text{det}(\boldsymbol{x},\tilde{\boldsymbol{\mu}})&:=\|\boldsymbol{x}-\tilde{\boldsymbol{\mu}}_\text{O1}\|^2-\epsilon-0.9\\
		h_\text{O2}^\text{det}(\boldsymbol{x},\tilde{\boldsymbol{\mu}})&:=\|\boldsymbol{x}-\tilde{\boldsymbol{\mu}}_\text{O2}\|^2-\epsilon-0.9.
		\end{align*}
Passing in between the obstacles O1 and O2 is not possibly due to the uncertainty in $\boldsymbol{X}$ and the risk predicates.
\end{example}

Increasing $c_m$ shrinks the set $\mathfrak{X}_m(c_m)$  so that Assumption~\ref{ass:3} (verifiable by Lemma \ref{lem:2})  poses a lower bound on $c_m$.  
\begin{theorem}\label{thm:3} 
	Let Assumption \ref{ass:3} hold and $\phi$ be an ReRiSITL formula in positive normal form. If $\boldsymbol{x}:\mathbb{R}_{\ge 0}\to\mathfrak{B}$ and $\boldsymbol{s}:\mathbb{R}_{\ge 0}\to\mathbb{B}^{|M^{\text{uc}}|}$ are such that $(\boldsymbol{x},\boldsymbol{s},\tilde{\boldsymbol{\mu}},t)\models \theta$, then it follows that $(\boldsymbol{x},\boldsymbol{s},\boldsymbol{X},t)\models \phi$.
	
	\begin{proof}
Due to Assumption \ref{ass:3}, $\boldsymbol{x}\in\mathfrak{X}_m(c_m)$ implies  $\boldsymbol{x}\in\mathfrak{X}_m^{\text{EV}}(\gamma_m)$, $\boldsymbol{x}\in\mathfrak{X}_m^{\text{VaR}}(\beta_m,\gamma_m)$, or $\boldsymbol{x}\in\mathfrak{X}_m^{\text{CVaR}}(\beta_m,\gamma_m)$ depending on the type of the predicate $m$. It is now straightforward to recursively show on the ReRiSITL semantics in Definition \ref{def:ReRiSITL} that  $(\boldsymbol{x},\boldsymbol{s},\tilde{\boldsymbol{\mu}},t)\models \theta$ implies $(\boldsymbol{x},\boldsymbol{s},\boldsymbol{X},t)\models \phi$ when $\boldsymbol{x}(t)\in\mathfrak{B}$, which holds by assumption. This follows since the semantics of ReRiSITL and ReSITL only differ on the predicate level and since negations are excluded since $\phi$ is in positive normal form. 
	\end{proof}
\end{theorem}

 An important task is  to pick the set of $c_m$. In general, we may induce conservatism  since the level sets of $\mathfrak{X}_m(c_m)$ may not be aligned with the level sets of  $\mathfrak{X}_m^{\text{EV}}(\gamma_m)$, $\mathfrak{X}_m^{\text{VaR}}(\beta_m,\gamma_m)$, and  $\mathfrak{X}_m^{\text{CVaR}}(\beta_m,\gamma_m)$. When linearity of $h_m(\boldsymbol{x},\boldsymbol{X})$ in $\boldsymbol{x}$ holds  as in Lemma \ref{lem:2},   conservatism can be avoided  \cite[Lemma~2]{lindemann2020control}.


If now, however, $\phi$ is not in positive normal form, there are two ways how to handle this case. The first way is to find $c_m$ for each $m\in\{1,\hdots,|M^{\text{Ri}}|\}$ according to \cite[Lemma~2]{lindemann2020control}, i.e., the set inclusion in Assumption \ref{ass:3} is replaced by an equality. More generally, a more elegant way is to bring $\phi$ into positive normal form, as for instance shown in \cite[Proposition 2]{sadraddini2015robust}. This would lead to a formula $\phi$ potentially having negations in front of some or all of the predicates, i.e., $\neg \mu_m^\text{Ri}$. For those predicates, we redefine the sets  $\mathfrak{X}_m^{\text{EV}}(\gamma_m)$, $\mathfrak{X}_m^{\text{VaR}}(\beta_m,\gamma_m)$, and $\mathfrak{X}_m^{\text{CVaR}}(\beta_m,\gamma_m)$ as 
\begin{align*}
\mathfrak{X}_m^{\text{EV}}(\gamma_m)&:=\{\boldsymbol{x}\in\mathfrak{B}|EV[-h_m(\boldsymbol{x},\boldsymbol{X})]> \gamma_m\}\\
\mathfrak{X}_m^{\text{VaR}}(\beta_m,\gamma_m)&:=\{\boldsymbol{x}\in\mathfrak{B}|VaR_{\beta_m}(-h_m(\boldsymbol{x},\boldsymbol{X}))> \gamma_m\}\\
\mathfrak{X}_m^{\text{CVaR}}(\beta_m,\gamma_m)&:=\{\boldsymbol{x}\in\mathfrak{B}|CVaR_{\beta_m}(-h_m(\boldsymbol{x},\boldsymbol{X}))> \gamma_m\}.
\end{align*}
 Note that only the sign of the inequality has changed compared to the definition in Section \ref{sec:const_sets}. For $c_m\in\mathbb{R}$, we then also redefine $\mathfrak{X}_m(c_m)$ as
\begin{align*}
\mathfrak{X}_m(c_m):=\{\boldsymbol{x}\in\mathfrak{B}|h_m(\boldsymbol{x},\tilde{\boldsymbol{\mu}})\le c_m\}.
\end{align*}
We would now again like to establish  the set inclusions as in Assumption \ref{ass:3} by a suitable choice of $c_m$ with these modified definitions. Note that these inclusions can then be similarly checked as in Lemma \ref{lem:2} (just reversing inequalities again).

\section{Reactive Planning Under ReSITL Specifications}
\label{sec:planning}

Following Section \ref{sec:determinization}, we can obtain an ReSITL formula $\theta$ from the ReRiSITL formula $\phi$. Motivated by the soundness result in Theorem \ref{thm:3}, we now  propose a reactive planning and control method that  leads to a satisfaction of the ReSITL formula $\theta$ that consequently leads to the satisfaction of the ReRiSITL formula $\phi$  (see also the top right box in Fig. \ref{fig:overview}).

In Section \ref{sec:abstraction}, we abstract the control system in \eqref{system_noise} into a timed signal transducer $TST_S$ (top left box  in Fig. \ref{fig:overview}). This abstraction is based on the assumption of existing logic-based feedback control laws from Section \ref{sec:control}. We then  modify $TST_{\theta}$ into $TST_{\theta}^\text{m}$ (bottom box  in Fig. \ref{fig:overview}), a product automaton between $TST_{\theta}$ and $TST_S$ that does not induce an exponential state explosion since $TST_{\theta}$ and $TST_S$ ``align'' in a suitable way due to the particular control laws in Section \ref{sec:control}.\footnote{$TST_{\theta}$ is a timed signal transducer for $\theta$ and constructed in the same way as $TST_{\phi}$ was obtained previously for $\phi$.} In Section \ref{sec:CEGIS}, we then present the reactive planning method that consists of a combination of a game-based approach and graph search techniques (boxes in the middle of  Fig. \ref{fig:overview}).

In Algorithm \ref{alg:3} presented below, we summarize the reactive planning algorithm that is presented in this section. In the remainder, we present and explain the steps of Algorithm~\ref{alg:3}. In line 1, abstract the ReSITL formula $\theta(\hat{M})$ into an MITL formula 
\begin{align*}
\varphi:=Tr(\theta(\hat{M}))=\theta(AP).
\end{align*} 
Note that we abstract $\theta(\hat{M})$, which depends on deterministic predicates and uncontrollable propositions $\hat{M}$ (recall that $\hat{M}:=M^{\text{det}} \cup M^{\text{uc}}$),  as opposed to $\phi(M)$ in Section \ref{sec:to_mitl} by the transformation $Tr(\cdot)$. Based on $\varphi$, construct 
\begin{align*}
TST_\varphi:=(S,s_{0},\Lambda,\Gamma,\boldsymbol{c},\iota,\Delta,\lambda,\gamma, \mathcal{A})
\end{align*}
 according to Section \ref{sec:mitl_to_ta} (Line 2 in Algorithm \ref{alg:3}). We again assume that uncontrollable propositions $p_i\in AP$, i.e., $p_i$ with $Tr^{-1}(p_i)\in \hat{M} \cap M^\text{uc}$, are modeled as in Fig. \ref{fig:uncontrollable_prop}. In Line 3, perform operations [O1] and [O2] on $TST_\varphi$\footnote{The notation in [O1] and [O2] needs to be slightly modified to account for $\theta$ instead of $\phi$. In particular,  $\boldsymbol{X}$  should be replaced with $\tilde{\boldsymbol{\mu}}$.}  to obtain the timed signal transducer 
 \begin{align*}
TST_\theta:=(S^\theta,s_{0},\Lambda,\Gamma,\boldsymbol{c},\iota,\Delta^\theta,\lambda,\gamma, \mathcal{A}^\theta).
 \end{align*}
 Note  that checking [O1] and [O2] is computationally  tractable if Assumption \ref{ass:1} holds due to the determinization in Section \ref{sec:determinization}.

\begin{algorithm}\label{alg:3}
	\caption{Reactive planning for ReSITL formula $\theta$.}
	\label{alg:3}
	\begin{algorithmic}[1]
		\State Obtain the MITL formula $\varphi:=Tr(\theta)$.
		\State Obtain $TST_\varphi$ according to Section \ref{sec:mitl_to_ta} and where uncontrollable propositions $p_i\in AP$, i.e., $p_i$ with $Tr^{-1}(p_i)\in \hat{M}  \cap M^\text{uc}$, are modeled as in Fig. \ref{fig:uncontrollable_prop}.
		\State Perform [O1] and [O2] to obtain $TST_\theta$.
		\State Obtain $TST_S$ according to Section \ref{sec:abstraction}.
		\State Perform [O3], [O4], and [O5] to obtain $TST_\theta^\text{m}$.
		\State Modify $TST_\theta^\text{m}$ to avoid Zeno behavior.
		\State Translate  $TST_\theta^\text{m}$ into ${RA}_C(TST_\theta^\text{m})$.
		\State Translate ${RA}_C(TST_\theta^\text{m})$ into $\overline{RA}_C(TST_\theta^\text{m})$.
		\State Run Algorithm \ref{alg:1} with the modified function $\bar{\pi}:2^{\overline{Q}}\to 2^{\overline{Q}}$ and $\overline{RA}_C(TST_\theta^\text{m})$ as the inputs to obtain $W$.
		\State Calculate the initial plan $d_\mu(t)$ based on $\overline{RA}_C(TST_\theta^\text{m})$ and obtain the associated control law $\boldsymbol{u}(\boldsymbol{x},t)$ (only possible if the conditions in Theorem \ref{thm:2} are satisfied).
		\While {$\boldsymbol{s}(t)=\boldsymbol{s}^\bot$}
		\If {$\boldsymbol{s}(t)\neq\boldsymbol{s}^\bot$}
		\State Recalculate $d_p(t)$ and $\boldsymbol{u}(\boldsymbol{x},t)$
		\EndIf
		\State Apply $\boldsymbol{u}(\boldsymbol{x},t)$ to \eqref{system_noise}
		\EndWhile
	\end{algorithmic}
\end{algorithm}

\subsection{Timed Abstraction of the Dynamical Control System}
\label{sec:abstraction}

 
In line 4 of Algorithm \ref{alg:3}, we abstract the system in \eqref{system_noise} into a timed signal transducer 
\begin{align*}
TST_S:=(\tilde{S},\tilde{S}_0,\tilde{\Lambda},\tilde{c},\tilde{\Delta},\tilde{\lambda}),
\end{align*}
see top left box  in Fig. \ref{fig:overview}. Note the absence of output labels, invariants, and a B\"uchi acceptance condition, and that $\tilde{c}$ is a scalar. The transition relation $\tilde{\Delta}$ is now based on the ability of the system to switch in finite time, by means of a feedback control law $\boldsymbol{u}_{\tilde{\delta}}(\boldsymbol{x},t)$  between elements in $Tr^{-1}(BC(TST_\theta))\subseteq BC(\tilde{\Lambda})$ where $\tilde{\Lambda}:=M$ and 
\begin{align*}
BC(TST_\theta):=\{z\in BC(AP)| \exists s \in S^\theta, \lambda(s)=z \}.
\end{align*}  
It is assumed that a library of such logic-based feedback control laws $\boldsymbol{u}_{\tilde{\delta}}(\boldsymbol{x},t)$ is available, e.g., as presented in Section \ref{sec:control}. Assume that $|\tilde{S}|= |Tr^{-1}(BC(TST_\theta))|$ and let $\tilde{\lambda}:\tilde{S}\to Tr^{-1}(BC(TST_\theta))$ where, for $\tilde{s}',\tilde{s}''\in \tilde{S}$ with $\tilde{s}'\neq \tilde{s}''$, it holds that $\tilde{\lambda}(\tilde{s}')\neq\tilde{\lambda}(\tilde{s}'')$ so that each state is uniquely labelled by $\tilde{\lambda}$, i.e., each state indicates exactly one Boolean formula from $Tr^{-1}(BC(TST_\theta))$. Note that $TST_\theta$ and $TST_S$ now ``align" in a way that will allow to avoid a state space explosion when forming a product automaton between them. A transition from $\tilde{s}$ to $\tilde{s}'$ is indicated by $(\tilde{s},\tilde{g},0, \tilde{s}')\in \tilde{\Delta}$ where $\tilde{g}$ is a guard that depends on \eqref{system_noise}. In particular, we assume that $\tilde{g}$ encodes intervals of the form $(C',C'')$, $[C',C'')$, $(C',C'']$, $[C',C'']$, or conjunctions of them, where $C',C''\in\mathbb{Q}_{\ge 0}$ with $C'\le C''$.
\begin{definition}[Transitions in $TST_S$]\label{def:transition_TST_S}
	There exists a transition $\tilde{\delta}:=(\tilde{s},\tilde{g},0, \tilde{s}')\in \tilde{\Delta}$ if, for all $\tau>0$ with $\tau\models \tilde{g}$ and for all $\boldsymbol{x}_0\in\mathbb{R}^n$ with $(\boldsymbol{x}_0,\boldsymbol{s}^\bot,\tilde{\boldsymbol{\mu}})\models \tilde{\lambda}(\tilde{s})$, there exists a control law $\boldsymbol{u}_{\tilde{\delta}}(\boldsymbol{x},t)$ so that the solution $\boldsymbol{x}(t)$ to \eqref{system_noise} is such that: 
	\begin{itemize}
		\item either, for all $t\in [0,\tau)$, $(\boldsymbol{x}(t),\boldsymbol{s}^\bot,\tilde{\boldsymbol{\mu}})\models \tilde{\lambda}(\tilde{s})$  and  $(\boldsymbol{x}(\tau),\boldsymbol{s}^\bot,\tilde{\boldsymbol{\mu}})\models \tilde{\lambda}(\tilde{s}')$
		\item or, for all $t\in [0,\tau]$, $(\boldsymbol{x}(t),\boldsymbol{s}^\bot,\tilde{\boldsymbol{\mu}})\models \tilde{\lambda}(\tilde{s})$ and there exists $\tau'>\tau$ such that, for all $t\in(\tau,\tau']$, $(\boldsymbol{x}(\tau'),\boldsymbol{s}^\bot,\tilde{\boldsymbol{\mu}})\models \tilde{\lambda}(\tilde{s}')$.
	\end{itemize}
	for which we define $\tilde{\lambda}(\tilde{\delta}):=\tilde{\lambda}(\tilde{s}')$ in the former and $\tilde{\lambda}(\tilde{\delta}):=\tilde{\lambda}(\tilde{s})$ in the latter case.
\end{definition}


The two types of transitions in the above definition can be thought of as transitioning into closed and open regions in $\mathbb{R}^n$, respectively.  Note that such a control law $\boldsymbol{u}_{\tilde{\delta}}(\boldsymbol{x},t)$ has to ensure invariance and finite-time reachability properties. Note also that $\boldsymbol{s}^\bot$ is used in Definition~\ref{def:transition_TST_S}  since controlled transitions will only happen when all uncontrollable propositions are false. Finally, the set $\tilde{S}_0$ consists of all elements $\tilde{s}_0\in \tilde{S}$ such that $(\boldsymbol{x}_0,\boldsymbol{s}^\bot,\tilde{\boldsymbol{\mu}})\models\tilde{\lambda}(\tilde{s}_0)$.

According to line 5 of Algorithm \ref{alg:3}, we next form a product automaton $TST_\theta^\text{m}$ (bottom box  in Fig. \ref{fig:overview}) of $TST_\theta$ and $TST_S$ that avoids a state space explosion that is typically the outcome of forming automata products. This follows since each input label of a state or transition in $TST_\theta$ corresponds to one state label of $TST_S$, i.e., $TST_\theta$ and $TST_S$ align in a way, so that $TST_\theta^\text{m}$ (defined below and corresponding to the product of $TST_\theta$ and $TST_S$) has no more states than $TST_\theta$.
Our approach relies on: 1) the removal of  transitions from $TST_\theta$, and 2) constraining guards $g$ of transitions in $TST_\theta$ to account for guards $\tilde{g}$ in $TST_S$. Let us, without loss of generality, assume that  each input label of a transition in $TST_\theta$ contains every literal from $\hat{M}$ and does not contain any disjunctions.\footnote{Note that each input label of a transition in $TST_\theta$ can be converted into full disjunctive normal form. Then, this transition can be split into several transitions, one for each disjunct, where each new input label corresponds to exactly one of the disjuncts.}
\begin{enumerate}
	\item[{[O3]}] For each transition $\delta:=(s,g,r,s')\in \Delta^\theta$ for which there exists $\boldsymbol{x}\in\mathbb{R}^n$ such that $(\boldsymbol{x},\boldsymbol{s}^\bot,\tilde{\boldsymbol{\mu}})\models Tr^{-1}(\lambda(\delta))$, remove $\delta$ if
	\begin{enumerate}
			\item there exists no transition $\tilde{\delta}:=(\tilde{s},\tilde{g},0,\tilde{s}')\in \tilde{\Delta}$  with $\lambda(s)=Tr(\tilde{\lambda}(\tilde{s}))$, and $\lambda(s')=Tr(\tilde{\lambda}(\tilde{s}'))$, and for which $(\boldsymbol{x},\boldsymbol{s}^\bot,\tilde{\boldsymbol{\mu}})\models\tilde{\lambda}(\tilde{\delta})$ implies $(\boldsymbol{x},\boldsymbol{s}^\bot,\tilde{\boldsymbol{\mu}})\models Tr^{-1}(\lambda(\delta))$.
	\end{enumerate} 
 Remove the corresponding $\delta$ from $\mathcal{A}^\theta$.
\end{enumerate}

We follow two goals with operation [O3]. First, we only consider to remove transitions that are induced by uncontrollable propositions being false, i.e., when $\boldsymbol{s}= \boldsymbol{s}^\bot$. This is important as we would like to keep transitions with $\boldsymbol{s}\neq \boldsymbol{s}^\bot$  for the reactive planning.  Note in particular that, if  there  exists $\boldsymbol{x}\in\mathbb{R}^n$ such that $(\boldsymbol{x},\boldsymbol{s}^\bot,\tilde{\boldsymbol{\mu}})\models Tr^{-1}(\lambda(\delta))$, then there exists no $\boldsymbol{x}\in\mathbb{R}^n$ such that $(\boldsymbol{x},\boldsymbol{s},\tilde{\boldsymbol{\mu}})\models Tr^{-1}(\lambda(\delta))$  for $\boldsymbol{s}\neq \boldsymbol{s}^\bot$. Second, we  remove such transitions if there exists no control law ${\boldsymbol{u}}_{\tilde{\delta}}$ that can simulate the transition in the system  \eqref{system_noise}. 


\begin{enumerate}
	\item[{[O4]}]  For  each transition $\delta_0:=(s_0,g,r,s')\in \Delta$, remove $\delta_0$ if $(\boldsymbol{x}_0,\boldsymbol{s}^\bot,\tilde{\boldsymbol{\mu}})\not\models Tr^{-1}(\lambda(s'))$ or if there exists no $\boldsymbol{s}\in \mathbb{B}^{|M^\text{uc}|}$ such that $(\boldsymbol{x}_0,\boldsymbol{s},\tilde{\boldsymbol{\mu}})\not\models Tr^{-1}(\lambda(\delta_0))$. Remove the corresponding $\delta_0$ from $\mathcal{A}^\theta$. 
\end{enumerate}

Operation [O4] takes care of the initial condition $\boldsymbol{x}_0$. If $s_0$ is removed in [O4], the problem is infeasible given the initial condition $\boldsymbol{x}_0$. 

Denote next the obtained sets by $S^\text{m}$, $\Delta^\text{m}$, and $\mathcal{A}^\text{m}$ for which  $S^\text{m}\subseteq S^\theta$, $\Delta^\text{m}\subseteq \Delta^\theta$, and $\mathcal{A}^\text{m}\subseteq \mathcal{A}^\theta$. We further take care of the timings including an additional clock into $TST_\theta$. Therefore, let $\boldsymbol{c}^\text{m}:=\begin{bmatrix}
\boldsymbol{c}^T & \tilde{c}\end{bmatrix}^T$ and perform the operation:
\begin{enumerate}
	\item[{[O5]}] For each transition $\delta^\text{m}:=(s,g,r,s')\in\Delta^\text{m}$ for which there exists $\boldsymbol{x}\in\mathbb{R}^n$ such that $(\boldsymbol{x},\boldsymbol{s}^\bot,\tilde{\boldsymbol{\mu}})\models Tr^{-1}(\lambda(\delta^\text{m}))$, let $g^\text{m}=g \wedge \tilde{g}$ where $\tilde{\delta}:=(\tilde{s},\tilde{g},0,\tilde{s}')\in\tilde{\Delta}$ with $\lambda(s)=Tr(\tilde{\lambda}(\tilde{s}))$, $\lambda(s')=Tr(\tilde{\lambda}(\tilde{s}'))$, and for which $(\boldsymbol{x},\boldsymbol{s}^\bot,\tilde{\boldsymbol{\mu}})\models\tilde{\lambda}(\tilde{\delta})$ implies $(\boldsymbol{x},\boldsymbol{s}^\bot,\tilde{\boldsymbol{\mu}})\models Tr^{-1}(\lambda(\delta))$. Replace $g$ and $r$ in $\delta^\text{m}$ with $g^\text{m}$ and $r^\text{m}$, respectively, where $r^\text{m}$ is obtained in an obvious manner. 
\end{enumerate}

We emphasize that adding $\tilde{c}$ and $\tilde{g}$ is crucial to ensure correctness. Let the modified timed signal transducer be denoted by 
\begin{align*}
TST^\text{m}_\theta:=(S^\text{m},s_{0},\Lambda,\Gamma,\boldsymbol{c}^\text{m},\iota,\Delta^\text{m},\lambda,\gamma, \mathcal{A}^\text{m})
\end{align*}
 and note that $L(TST^\text{m}_\theta)\subseteq L(TST_\theta)\subseteq L(TST_\varphi)$.
\begin{remark}
	The operations [O3]-[O5] result in the timed signal transducer $TST^\text{m}_\theta$ that, by construction, restricts the behavior of $TST_\theta$ exactly to the behavior allowed by $TST_S$ and corresponds hence to a product automaton without exhibiting an exponential state space explosion.
\end{remark}

\subsection{Reactive Plan Synthesis}
\label{sec:CEGIS}

\begin{figure}
	\centering
	\includegraphics[scale=0.4]{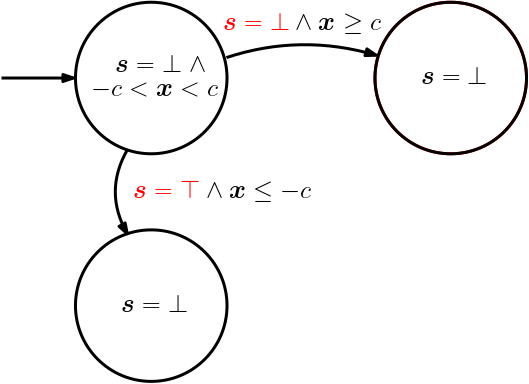}\caption{Illustration of Example \ref{ex:1} and why a modified definition of $\pi(W)$ is needed to avoid discontinuities in $\boldsymbol{x}(t)$.}\label{fig:cont_replan}
\end{figure}

Based on $TST^\text{m}_\theta$, let us now present the reactive planning method depicted in the boxes in the middle  of Fig. \ref{fig:overview}. We first derive a \emph{nominal plan} $d_\mu:\mathbb{R}_{\ge 0}\to BC(M)$ from $TST^\text{m}_\theta$ based on the assumption that $\boldsymbol{s}(t)=\boldsymbol{s}^\bot$ for all $t\in\mathbb{R}_{\ge 0 }$. This plan is executed until  $\boldsymbol{s}(t_\text{replan})\neq\boldsymbol{s}^\bot$ for some $t_\text{replan}\in\mathbb{R}_{\ge 0}$, the moment when \emph{reactive and online replanning} is needed. In line 6 of Algorithm \ref{alg:3}, let $TST^\text{m}_\theta$ again be modified to not exhibit Zeno behavior and let 
\begin{align*}
\overline{RA}_C(TST^\text{m}_\theta):=(\overline{Q},\overline{q}_0,\overline{\Delta}_R,\overline{\mathcal{A}}_R)
\end{align*}
 be the region automaton of $TST^\text{m}_\theta$ based on $(S^\text{m}\times \mathbb{R}_{\ge 0}^O,\Rightarrow_C)$ and Definitions \ref{def:trans_rel_} and \ref{def:region_} (lines 7 and 8 of Algorithm \ref{alg:3}).\footnote{ Definitions \ref{def:trans_rel_} and \ref{def:region_} need to be altered to account for using $TST^\text{m}_\theta$ instead of $TST_\phi$ in an obvious manner.} Replanning may now require to take, at an unknown time instant $t_\text{replan}$, a transition that is not contained within the nominal plan. Those instances may possibly require an infeasible discontinuity in the physical state $\boldsymbol{x}$ that we need to rule out. 
\begin{example}\label{ex:1}
	To illustrate the aforementioned issue, consider  Fig. \ref{fig:cont_replan}. For the top left state, there exist two transitions to the top right and the bottom left state.  Assume the former transition can be realized by the control law $\boldsymbol{u}_{\tilde{\delta}}(\boldsymbol{x},t)$. Starting from the top left state, the initial plan will consider the transition with $\boldsymbol{s}=\bot$ to the top right state implying that $\boldsymbol{u}_{\tilde{\delta}}(\boldsymbol{x},t)$ is used until time $t=t_\text{replan}$ such that  $0<\boldsymbol{x}(t_\text{replan})<c$. After replanning, however, the other transition with $\boldsymbol{s}=\top$ to the bottom left state has to instantaneously be taken requiring to immediately achieve $\boldsymbol{x}\le -c$. Such a discontinuity in $\boldsymbol{x}(t)$ is  not realizable in \eqref{system_noise} that only admits continuous $\boldsymbol{x}(t)$. 
\end{example} 

One way of dealing with this  issue is to modify the predecessor operator. Recall therefore that a state $\overline{q}\in\overline{Q}$ in $\overline{RA}_C(TST^\text{m}_\theta)$ consists of the elements $\bar{q}:=(s,\alpha,i)\in Q\subseteq S^\text{m}\times A \times \{1,\hdots,|\mathcal{A}^\text{m}|\}$ and redefine now $\pi(W)$ to 
\begin{align*}
\hat{\pi}(W)&:=\{ \overline{q}\in \overline{Q}| \forall \boldsymbol{s}\in \mathbb{B}^{|M^{\text{uc}}|},  \exists  (\overline{q},\delta,\overline{q}')\in\overline{\Delta}_R \text{ s.t. }\text{ 1) }  \overline{q}'\in W, \text{and 2) } \forall\boldsymbol{x}\in\mathbb{R}^n \text{ s.t. } \\
&(\boldsymbol{x},\boldsymbol{s}^\bot,\tilde{\boldsymbol{\mu}})\models Tr^{-1}(\lambda(s)),(\boldsymbol{x},\boldsymbol{s},\tilde{\boldsymbol{\mu}})\models Tr^{-1}(\lambda(\delta))  \text{ and }   
(\boldsymbol{x},\boldsymbol{s}^\bot,\tilde{\boldsymbol{\mu}})\models Tr^{-1}(\lambda(s')) 
\}
\end{align*}

The second condition in $\hat{\pi}(W)$ now additionally ensures that all $\boldsymbol{x}$ that satisfy the state label of $s$ also satisfy the state labels of the transition $\delta$ as well as the next state $s'$.  As a consequence, an instantaneous transition from $\overline{q}$ to $\overline{q}'$ due to $\boldsymbol{s}(t)\neq \boldsymbol{s}^\bot$ can happen without requiring that $\boldsymbol{x}(t)$ is discontinuous. We emphasize, again, that this condition is necessary with respect to the solutions to \eqref{system_noise}. Let $W$ be obtained from Algorithm \ref{alg:1} with $\overline{RA}_C(TST^\text{m}_\theta)$ and $\hat{\pi}:2^{\overline{Q}}\to 2^{\overline{Q}}$ as the input (line 9 of Algorithm \ref{alg:3}). 


\subsubsection{Initial Plan Synthesis:} For line 10 in Algorithm \ref{alg:3}, let $d_p(t)$, as opposed to Section \ref{sec:tst_to_reg}, now be obtained from $\overline{RA}_C(TST^\text{m}_\theta)$ as follows. We find, using graph search techniques, a sequence $\mathfrak{q}:=(\overline{q}_0,\overline{q}_1,\hdots):=(\mathfrak{q}_p,\mathfrak{q}_s^\omega)$ satisfying the B\"uchi acceptance condition $\overline{\mathcal{A}}_R$   with 
\begin{align*}
\overline{q}_j\in\overline{Q}\cap W
\end{align*}
 for each $j\in \mathbb{N}$ and where $(\overline{q}_j,\delta_{t,j},\overline{q}_{j+1})\in \overline{\Delta}_R$ so that, for each $\delta_{t,j}$,  there exists $\boldsymbol{x}\in\mathbb{R}^n$ such that $(\boldsymbol{x},\boldsymbol{s}^\bot,\tilde{\boldsymbol{\mu}})\models Tr^{-1}(\delta_{t,j})$. Note in particular the intersection with $W$ that will ensure that replanning is possible whenever $\boldsymbol{s}(t_\text{replan})\neq\boldsymbol{s}^\bot$ for some $t_\text{replan}\in\mathbb{R}_{\ge 0}$, as elaborated on in the next section.  Additionally and for the initial transition $\delta_0$, we again require that $\gamma(\delta_{0})=y$ to indicate $(\boldsymbol{x},\boldsymbol{s},\tilde{\boldsymbol{\mu}},0)\models\theta$. Note in particular the restriction to $\overline{q}_j\in\overline{Q}\cap W$ which will allow to replan if $\boldsymbol{s}(t_\text{replan})\neq \boldsymbol{s}^\bot$ for some $t_\text{replan}\in\mathbb{R}_{\ge 0}$. We again find timings $\bar{\tau}:=(\tau_0,\tau_1,\hdots):=(\bar{\tau}_p,\bar{\tau}_s^\omega)$  that are associated with  $\mathfrak{q}$. Such a plan $d_p(t)$ is guaranteed to exist if the conditions in Theorem \ref{thm:2} are satisfied. Recall that $T_j:=\sum_{k=0}^j \tau_j$,  and define
\begin{align}\label{eq:plan11333}
d_p(t):=\begin{cases}
\lambda(\delta_{t,j}) &\text{if } t= T_j\\ 
\lambda(s_j) &\text{if }  T_j < t <T_{j+1}
\end{cases}
\end{align}


  We can now define the control law $\boldsymbol{u}(\boldsymbol{x},\boldsymbol{s},t)$ based on the plan $d_\mu(t):=Tr^{-1}(d_p(t))$. Recall therefore that each transition $\delta_{t,j}$ is associated, when projecting back to $TST_S$, with a control law $\boldsymbol{u}_{\tilde{\delta}_{t,j}}(\boldsymbol{x},t)$ as explained in Section \ref{sec:abstraction}. Recall the definition of $T_j$ and let
\begin{align*}
	\boldsymbol{u}(\boldsymbol{x},\boldsymbol{s},t):=\begin{cases}\boldsymbol{u}_{\tilde{\delta}_{1}}(\boldsymbol{x},t) &\hspace{-0.2cm}\text{for  } t\in[0,T_1)\\
	\boldsymbol{u}_{\tilde{\delta}_{j+1}}(\boldsymbol{x},t-T_j) &\hspace{-0.2cm}\text{for  } t\in(T_j,T_{j+1}) \text{ with } j\ge 2
	\end{cases}
\end{align*}
and, for $t=T_j$ with $j\ge 2$, let
\begin{align*}
	\boldsymbol{u}(\boldsymbol{x},\boldsymbol{s},T_j):=\begin{cases}
		\boldsymbol{u}_{\tilde{\delta}_{j+1}}(\boldsymbol{x},0) &\text{ if } \tilde{\lambda}(\tilde{s}_{j+1})= d_\mu(T_j)\\
		\boldsymbol{u}_{\tilde{\delta}_{j}}(\boldsymbol{x},\tau_j) &\text{ if } \tilde{\lambda}(\tilde{s}_{j})= d_\mu(T_j).	
	\end{cases}
\end{align*}
Note that $\boldsymbol{u}(\boldsymbol{x},\boldsymbol{s},T_j)$ in particular accounts for the two types of transitions in Definition \ref{def:transition_TST_S}.

\begin{corollary}\label{cor:3}
	Assume that $\boldsymbol{s}(t)=\boldsymbol{s}^\bot$ for all $t\in\mathbb{R}_{\ge 0}$, $\overline{q}_0\in W$, and there exists $(\overline{q}_0,\delta_0,\overline{q}')\in\Delta_R$ with $\gamma(\delta_0)=y$, then $d_p(t)$ as in \eqref{eq:plan11333} exists and $\boldsymbol{u}(\boldsymbol{x},\boldsymbol{s},t)$  results in $(\boldsymbol{x},\boldsymbol{s},\tilde{\boldsymbol{\mu}},0)\models \theta$.
	
	\begin{proof}
	Similar to Theorem \ref{thm:2} and by the construction of $TST^\text{m}_\theta$, it follows that $\theta$ is satisfiable given that $\overline{q}_0\in W$ and that there exists $(\overline{q}_0,\delta_0,\overline{q}')\in\Delta_R$ with $\gamma(\delta_0)=y$. It directly follows  that, in this case, a plan $d_p(t)$ exists. Note next that by construction of $TST_S$ and $TST^\text{m}_\theta$, each transition $\delta$ in $TST^\text{m}_\theta$ can be realized in \eqref{system_noise} by an associated control law $\boldsymbol{u}_{\tilde{\delta}}(\boldsymbol{x},t)$. By the construction of the plan $d_p(t)$ and the associated control law $\boldsymbol{u}(\boldsymbol{x},t)$, it follows trivially that $\boldsymbol{u}(\boldsymbol{x},\boldsymbol{s},t)$, build from a sequence of such $\boldsymbol{u}_{\tilde{\delta}}(\boldsymbol{x},t)$,  results in $(\boldsymbol{x},\boldsymbol{s},\tilde{\boldsymbol{\mu}},0)\models \theta$.
	\end{proof}
\end{corollary}

\subsubsection{Reactive and online replanning: } If hence $\boldsymbol{s}(t)=\boldsymbol{s}^\bot$ for all $t\in\mathbb{R}_{\ge 0}$, there is nothing left to do and we apply $\boldsymbol{u}(\boldsymbol{x},\boldsymbol{s},t)$ as in line 14 of Algorithm \ref{alg:3}. If, however, $\boldsymbol{s}(t_\text{replan})\neq\boldsymbol{s}^\bot$ for some $t_\text{replan}\in\mathbb{R}_{\ge 0}$, we need to replan and update our plan $d_\mu(t)$ that may be violated by this particular $\boldsymbol{s}(t_\text{replan})$ (lines 12 and 13 in Algorithm \ref{alg:3}). Assume that, at time $t_\text{replan}$, the system is in state $\overline{q}_{j^*}$.  We then find an updated sequence $\mathfrak{q}_\text{replan}:=(\overline{q}_{j^*},\overline{q}_{j^*+1},\hdots)$ satisfying the B\"uchi acceptance condition $\overline{\mathcal{A}}_R$ again with 
\begin{align*}
\overline{q}_j\in\overline{Q}\cap W
\end{align*}
 for each $j>j^*$  and where  $(\overline{q}_j,\delta_{t,j},\overline{q}_{j^*+1})\in \overline{\Delta}_R$ so that 1) $(\boldsymbol{x}(t_\text{replan}),\boldsymbol{s}(t_\text{replan}),\tilde{\boldsymbol{\mu}})\models Tr^{-1}(\delta_{t,j^*})$, and 2) for each $\delta_{t,j}$ with $j>j^*$ there exists $\boldsymbol{x}\in\mathbb{R}^n$ such that $(\boldsymbol{x},\boldsymbol{s}^\bot,\tilde{\boldsymbol{\mu}})\models Tr^{-1}(\delta_{t,j})$. If $t_\text{replan}=0$, it is additionally required that $\gamma(\delta_{t,j^*})=y$. We again find timings $\bar{\tau}:=(\tau_{j^*},\tau_{j^*+1},\hdots)$  that are associated with  $\mathfrak{q}_\text{replan}$. Based on this updated sequence, we recalculate $d_p(t)$ in \eqref{eq:plan11333} and $\boldsymbol{u}(\boldsymbol{x},\boldsymbol{s},t)$ in an obvious manner. 

\begin{theorem}\label{thm:4}
	Assume that $\boldsymbol{s}(t)$ is according to Assumption \ref{ass:2}, $\overline{q}_0\in W$, and there exists $(\overline{q}_0,\delta_0,\overline{q}')\in\Delta_R$ with $\gamma(\delta_0)=y$, then finding an initial plan $d_p(t)$ and updating  $\boldsymbol{u}(\boldsymbol{x},\boldsymbol{s},t)$ in the previously described manner in case that $\boldsymbol{s}(t_\text{replan})\neq \boldsymbol{s}^\bot$  results in $(\boldsymbol{x},\boldsymbol{s},\tilde{\boldsymbol{\mu}},0)\models \theta$.
	
	\begin{proof}
		The assumptions that $\overline{q}_0\in W$ and that there exists $(\overline{q}_0,\delta_0,\overline{q}')\in\Delta_R$ with $\gamma(\delta_0)=y$, again guarantee that there exists an initial plan $d_p(t)$. Due to the properties of $W$ and given that $\boldsymbol{s}(t)$ is according to Assumption \ref{ass:2}, it holds that a new plan and an updated $\boldsymbol{u}(\boldsymbol{x},\boldsymbol{s},t)$ can always be found whenever $\boldsymbol{s}(t_\text{replan})\neq \boldsymbol{s}^\bot$. Each such instantaneous transition is well defined in the sense of not requiring a discontinuity in $\boldsymbol{x}(t)$ due to the modified definition of $\pi(W)$. 
		\end{proof}
\end{theorem}

We remark that Assumption \ref{ass:2} is not only necessary for the game-based approach in Algorithm \ref{alg:1}, but that the assumption is also necessary to be able to replan. Without Assumption \ref{ass:2}, there is no information about the value of $\boldsymbol{s}(t)$ shortly after $t_\text{replan}$.  By Assumption \ref{ass:2}, there  follows an open time interval in which $\boldsymbol{s}(t)=\boldsymbol{s}^\bot$ after $t_\text{replan}$ so that a next state can be selected whose state label is satisfied by $\boldsymbol{s}^\bot$. Further note that Assumption \ref{ass:2} effectively poses an upper bound on the frequency of times that replanning is initiated. 

To conclude this section, we note that a combination of graph search techniques and a game-based approach has been presented. The game-based approach ensures that it is always possible to make progress towards satisfying the B\"uchi acceptance condition by ruling out `bad' transitions, while graph search techniques actually enforce this progress.

\subsection{Feedback Control under STL Specifications}
\label{sec:control}

In this section, we discuss the control laws $\boldsymbol{u}_{\tilde{\delta}}(\boldsymbol{x},t)$ that are supposed to achieve the transitions $\tilde{\delta}:=(\tilde{s},\tilde{g},0,\tilde{s}')\in\tilde{\Delta}$ in Definition \ref{def:transition_TST_S} for the timed abstraction $TST_S$. In particular, such transitions can be captured by the STL formulas
\begin{align}
G_{[0,\tau)}\mu_\text{inv}(\boldsymbol{x}) \wedge F_{\tau} \mu_\text{reach}(\boldsymbol{x})\wedge G_{[0,\tau]}\mu_\text{ws}(\boldsymbol{x}),\label{eq:transition_until} \\
G_{[0,\tau]}\mu_\text{inv}(\boldsymbol{x}) \wedge G_{(\tau,\tau']} \mu_\text{reach}(\boldsymbol{x})\wedge G_{[0,\tau']}\mu_\text{ws}(\boldsymbol{x})\label{eq:transition_until_}
\end{align}
where $\mu_\text{inv}(\boldsymbol{x}):= \tilde{\lambda}(\tilde{s})$ and $\mu_\text{reach}(\boldsymbol{x}):=\tilde{\lambda}(\tilde{s}')$ are deterministic predicates as in \eqref{eq:STL_predicate} and where $\tau\in\mathbb{R}_{> 0}$ with $\tau\models\tilde{g}$, while $\mu_\text{ws}(\boldsymbol{x}) $ encodes a compact set $\mathfrak{B}$ according to Section \ref{sec:determinization}; $\mathfrak{B}$ can be any compact set, typically the workspace. With $\mu_\text{inv}(\boldsymbol{x})$,  $\mu_\text{reach}(\boldsymbol{x})$, and $\mu_\text{ws}(\boldsymbol{x})$, we can now associate predicate functions $h_\text{inv}(\boldsymbol{x})$, $h_\text{reach}(\boldsymbol{x})$, and $h_\text{ws}(\boldsymbol{x})$. 

There is a plethora of recent works that have addressed the problem  of controlling systems as in \eqref{system_noise} under spatio-temporal constraints as in \eqref{eq:transition_until} or \eqref{eq:transition_until_}. In particular \cite{lindemann2018control,garg2019control} address the control problem by time-varying  control-barrier functions and fixed time control Lyapunov functions, respectively. For robotic specific problem setups,  funnel control laws to solve \eqref{eq:transition_until} or \eqref{eq:transition_until_} have also appeared in \cite{verginis2018timed}, while optimization-based methods are presented in \cite{fiaz2019fast}. Another approach, relying on time-varying vector fields, has appeared in \cite{mavridis2019robot}. We are, purposefully and with respect to page limitations, not presenting a specific type of feedback control law here and emphasize that our proposed reactive planning method is agnostic to feedback control laws that can achieve the STL specification as in \eqref{eq:transition_until} or \eqref{eq:transition_until_}. Note that the previously mentioned works pose certain assumptions on the systems dynamics in \eqref{system_noise} as well as on the form of $h_\text{inv}(\boldsymbol{x})$, $h_\text{reach}(\boldsymbol{x})$, and $h_\text{ws}(\boldsymbol{x})$. We remark that controlling systems under timed specifications of the type in \eqref{eq:transition_until} or \eqref{eq:transition_until_} has recently attracted interest in the research community so that we expect more progress in this respect.

\section{Completeness and Complexity}
In summary, the presented framework consists of: 1) translating the ReRiSITL specification $\phi$ into a ReSITL specification $\theta$ in Section \ref{sec:determinization}, and 2) reactive planning under this ReSITL specifications $\theta$ in Section \ref{sec:planning}, as summarized in Algorithm \ref{alg:3}. The framework is sound in the sense of Theorems \ref{thm:3}  and \ref{thm:4}, but not necessarily complete, i.e., there may exist a solution even though we may not find it. There are three reasons for such conservatism.  First, the translation from the ReRiSITL specification $\phi$ to the ReSITL specification $\theta$ may induce conservatism as discussed in Section \ref{sec:determinization}. Second, in line 6 of Algorithm \ref{alg:3}, we need to modify $TST_\theta^\text{m}$ to avoid Zeno behavior. This operation potentially induces conservatism that can, however, be reduced as also discussed previously. Third,  the construction of nonlinear control laws, presented in Section \ref{sec:control}, may introduce conservatism. This is inherent in nonlinear control and we do not view this as a drawback of our method.

The presented framework consists of several computationally expensive operations. Fortunately, these operations can be performed offline. We focus on space complexity. First, the translation from the MITL formula $\varphi$ to the timed signal transducer $TST_\varphi$ induces $O(|\varphi|M)$ clocks and $2^{O(|\varphi|M)}$ states where $|\varphi|$ denotes the complexity of $\varphi$ and $M$ is related to the  length of the maximum time interval  in $\varphi$ (see  \cite[Theorem 6.7]{ferrere2019real}).  Operations [O1] and [O2], which transform $TST_\varphi$ into $TST_\theta$, ease the complexity by removing a considerable  number of states and transitions from $TST_\varphi$. An exact number is in general not quantifiable as those removals depend on  predicate dependencies in the specification $\theta$. Operations [O3] and [O4] further remove states and transitions from $TST_\theta$ to obtain the product automaton $TST_\theta^\text{m}$. Note that we obtain computational benefits over existing methods that would induce additional $O(|S^\theta||\tilde{S}|)$ states.  The operation $RA_C(TST_\theta^\text{m})$ results in an automaton with $O(|S^\text{m}|\text{len}(\boldsymbol{c}_m))$ states where $\text{len}(\boldsymbol{c}_m)$ denotes the length of clock constraints in  $TST_\theta^\text{m}$ (see  \cite[Section 4.3]{alur1994theory}). The translation from $RA_C(TST_\theta^\text{m})$ to $\overline{RA}_C(TST_\theta^\text{m})$ results in an automaton with  $O(|Q||\mathcal{A}_R|)$ states, which can considerably be reduced as discussed in Remark \ref{rem:5555}. The time complexity of Algorithm \ref{alg:1} and graph search techniques to find a plan $d_p(t)$ follows standard arguments. Operations [O1], [O2], [O3], and [O4] involve solving  nonlinear mixed integer programs, and in particular mixed integer linear programs when Assumption \ref{ass:1} holds.

\section{Simulations}
\label{sec:simulations}

We consider a unicycle model with dynamics 
\begin{align*}
\dot{\boldsymbol{z}}=f(\boldsymbol{z})+g(\boldsymbol{z})\boldsymbol{u}
\end{align*}
and where the state is given as
\begin{align*}
\boldsymbol{z}:=\begin{bmatrix} \boldsymbol{x}^T & x_a\end{bmatrix}^T:=\begin{bmatrix} x_x & x_y & x_a\end{bmatrix}^T
\end{align*}
 to model the two-dimensional position and orientation, respectively. Here, 
 \begin{align*}
 \boldsymbol{u}:=\begin{bmatrix}v & \omega \end{bmatrix}^T
 \end{align*} 
 contains the translational and rotational control inputs. In particular, let 
 \begin{align*}
 f(\boldsymbol{x}):=0.5\cdot\begin{bmatrix} -\text{sat}(x_x) & -\text{sat}(x_y) & 0 \end{bmatrix}^T
 \end{align*}
  where $\text{sat}(x)= x$ if $|x|\le 1$ and $\text{sat}(x)=1$ otherwise. Furthermore, let 
  \begin{align*}
  g(\boldsymbol{x}):=\begin{bmatrix} \cos(x_a) & 0 \\ \sin(x_a) & 0 \\ 0 & 1 \end{bmatrix}.
  \end{align*}
   To obtain $\boldsymbol{u}(\boldsymbol{z},t)$, we use here the time-varying control barrier functions from \cite{lindemann2018control}. In particular, time-varying control barrier functions adapted for nonholonmic systems from \cite{lindemann2020control} are used for which no knowledge of $f(\boldsymbol{z})$ is required.

For this system, the imposed ReRiSITL specification $\phi$ is the one given in Example \ref{ex:2}. The specification $\phi$ is rich enough to illustrate all theoretical findings (i.e., how to deal with risk predicates, uncontrollable propositions, and past temporal operators) and yet basic enough to explain all subtleties of $\phi$ and the reactive and risk-aware control sythesis. 

Recall the determinization of risk predicates according to Section \ref{sec:determinization} in Example \ref{ex:3} resulting in the ReSITL specification
\begin{align*}
\theta&:=Tr(\phi)= F_{(0,5)} \mu_\text{R1}^\text{det}\wedge G_{[0,\infty)}\Big(\mu_\text{O1}^\text{det}\wedge \mu_\text{O2}^\text{det} \wedge \big(\underline{F}_{(0,1)}\mu^\text{uc}\implies F_{(0,3)}\mu_\text{R2}^\text{det}\big)\Big).
\end{align*}
for which initially $(\boldsymbol{x}(0),\boldsymbol{s}^\bot,\tilde{\boldsymbol{\mu}})\models \neg\mu_\text{R1}^\text{det}\wedge \mu_\text{O1}^\text{det}\wedge \mu_\text{O2}^\text{det} \wedge \neg \mu_\text{R2}^\text{det}$ is assumed. For the construction of $TST_S$ in Section \ref{sec:abstraction}, we assume that we have control laws $\boldsymbol{u}_{\tilde{\delta}}(\boldsymbol{x},t)$ that can accomplish each transition $\tilde{\delta}$ as per Definition \ref{def:transition_TST_S} with $\tilde{g}:=[1,\infty)$. 

\emph{Setting 1:} With respect to Assumption \ref{ass:2}, we first assume that $\zeta:=1$. Recall that $\zeta$ determines the frequency by which the uncontrollable event $\mu^\text{uc}$ may occur. In this case, the set $W$ does not contain the element $\overline{q}_0$, i.e., $\overline{q}_0\not\in W$, so that by Theorem  \ref{thm:4} no plan $d_\mu(t)$ is found that satisfies $\theta$ and consequently $\phi$. Note that this follows mainly since $\boldsymbol{s}(t)=\text{proj}_{\mu^\text{uc}}(s)(t)=\top$ may occur within $\zeta$ time unit intervals implying that, in the worst case, $\mu_\text{R2}^\text{Ch}$ should always be true so that there is no time to satisfy $\mu_\text{R1}^\text{Ch}$.

\begin{figure}
	\centering
	\includegraphics[scale=0.5]{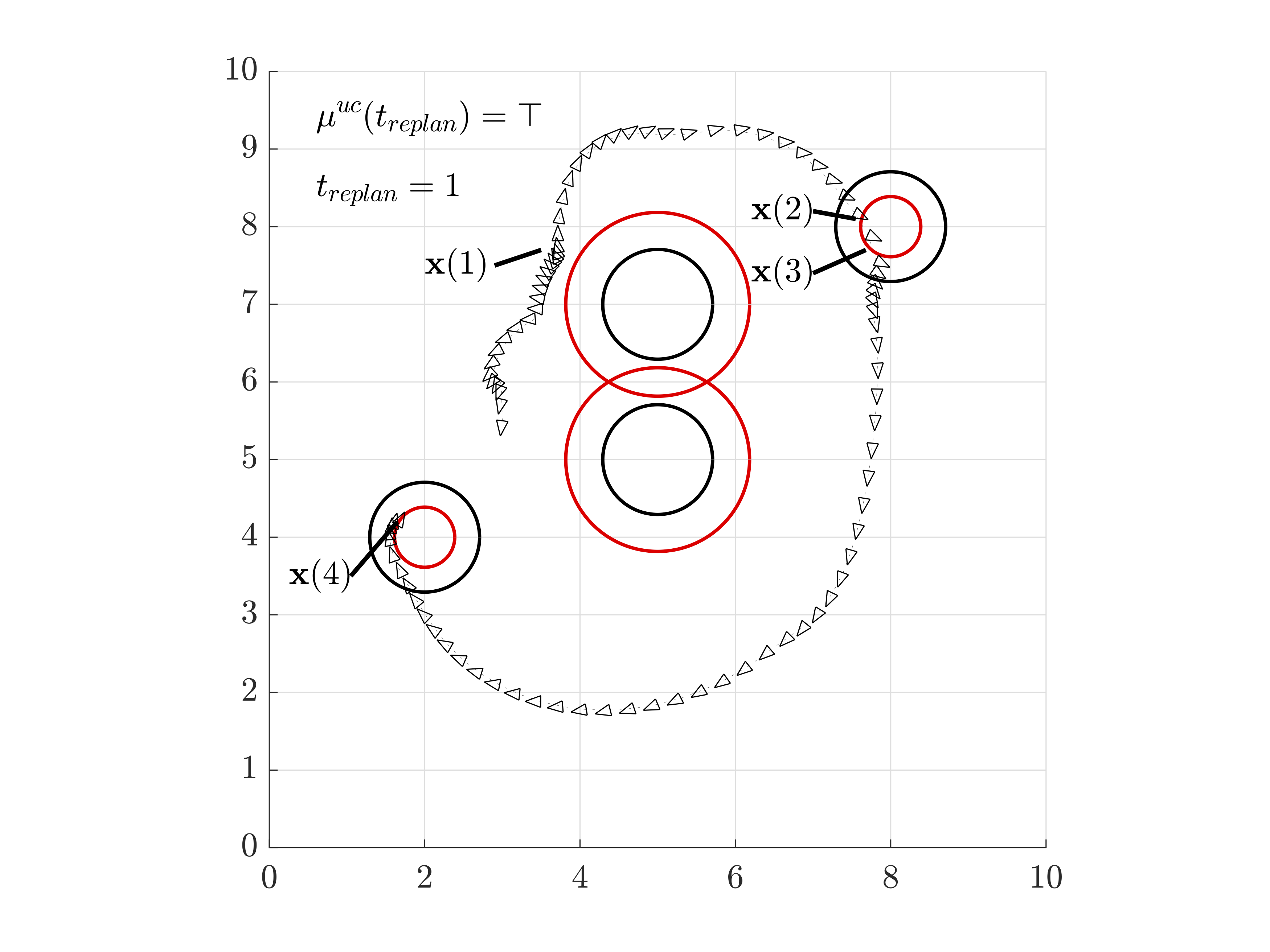}\caption{Unicycle model  for the ReRiSTL specification $\phi$ with $\zeta:=5$ and when an uncontrollable event occurs.}\label{fig:sim_res}
\end{figure}

\emph{Setting 2:} By increasing $\zeta$, the frequency by which the uncontrollable event $\mu^\text{uc}$ may occur is decreased. We set $\zeta:=5$ and now observe that $\overline{q}_0\in W$. The synthesized initial plan $d_\mu(t)$ is as follows.
\begin{align*}
d_\mu(t):=\begin{cases}
\neg \mu_\text{R1}^\text{det} \wedge \mu_\text{O1}^\text{det}\wedge \mu_\text{O2}^\text{det} \wedge \neg \mu_\text{R2}^\text{det}\wedge \neg\mu^\text{uc} & t\in(0,4)\\
\mu_\text{R1}^\text{det} \wedge \mu_\text{O1}^\text{det}\wedge \mu_\text{O2}^\text{det} \wedge \neg \mu_\text{R2}^\text{det}\wedge \neg\mu^\text{uc} & t\in[4,5.7]\\
\neg \mu_\text{R1}^\text{det} \wedge \mu_\text{O1}^\text{det}\wedge \mu_\text{O2}^\text{det} \wedge \neg \mu_\text{R2}^\text{det}\wedge \neg\mu^\text{uc} & t\in(5.7,\infty).
\end{cases}
\end{align*}
However, now assume that $\boldsymbol{s}(1)=\text{proj}_{\mu^\text{uc}}(1)=\top$ so that at $t_\text{replan}=1$ replanning is needed. Our revised plan then is
\begin{align*}
d_\mu(t):=\begin{cases}
\neg \mu_\text{R1}^\text{det} \wedge \mu_\text{O1}^\text{det}\wedge \mu_\text{O2}^\text{det} \wedge \neg \mu_\text{R2}^\text{det}\wedge \neg\mu^\text{uc} & t\in(0,1)\\
\neg \mu_\text{R1}^\text{det} \wedge \mu_\text{O1}^\text{det}\wedge \mu_\text{O2}^\text{det} \wedge \neg \mu_\text{R2}^\text{det}\wedge \mu^\text{uc} & t=1\\
\neg \mu_\text{R1}^\text{det} \wedge \mu_\text{O1}^\text{det}\wedge \mu_\text{O2}^\text{det} \wedge \neg \mu_\text{R2}^\text{det}\wedge \neg\mu^\text{uc} & t\in(1,2)\\
 \mu_\text{R1}^\text{det} \wedge \mu_\text{O1}^\text{det}\wedge \mu_\text{O2}^\text{det} \wedge \neg \mu_\text{R2}^\text{det}\wedge \neg\mu^\text{uc} & t\in[2,3)\\
  \neg\mu_\text{R1}^\text{det} \wedge \mu_\text{O1}^\text{det}\wedge \mu_\text{O2}^\text{det} \wedge \neg \mu_\text{R2}^\text{det}\wedge \neg\mu^\text{uc} & t\in[3,4)\\
    \neg\mu_\text{R1}^\text{det} \wedge \mu_\text{O1}^\text{det}\wedge \mu_\text{O2}^\text{det} \wedge \mu_\text{R2}^\text{det}\wedge \neg\mu^\text{uc} & t\in[4,\infty),\\
\end{cases}
\end{align*}
i.e., to prepone satisfying $\mu_\text{R1}^\text{det}$ and to satisfy $\mu_\text{R2}^\text{det}$ right after and within $3$ time units from when $\mu^\text{uc}$ happened. The simulation results for this case are depicted in Fig. \ref{fig:sim_res}. 

Simulations were performed on a 1.4 GHz quad-core Intel Core i5 with 8 GB RAM. Construction of $TST_\theta$ and $\overline{RA}_C(TST_\theta)$ took $2.5$ s and $78.5$ s, respectively, while Algorithm \ref{alg:1} and the graph search took $130$ s and $15.5$ s, respectively. All implementations are made in MATLAB, without optimizing for performance, and can be found under  \cite{dropbox}. A short animation can also be found in \cite{dropbox}.

\section{Conclusion}
\label{sec:conclusion}

This paper has presented reactive risk signal temporal logic (ReRiSTL) as a significant extension of signal temporal logic (STL). ReRiSTL  additionally allows to consider the risk of not satisfying an ReRiSTL specification as well as allowing to consider environmental events such as sensor failures. We have then proposed an algorithm to check if such an ReRiSTL specification is satisfiable. Lastly, we have proposed a reactive planning and control framework for dynamical systems under ReRiSTL specifications by combining a game-based approach with graph search techniques.

\section*{Acknowledgment}

The authors would like to thank Professor Antoine Girard for providing useful feedback in stating  Defintion \ref{def:satisfiability}. 

\bibliographystyle{IEEEtran}
\bibliography{literature}

\addtolength{\textheight}{-12cm}   

\end{document}